\documentclass[twocolumn,trackchanges]{aastex701}

\usepackage{graphicx}	
\usepackage{amsmath}	
\usepackage{amssymb}
\usepackage{siunitx} 
\usepackage{textcomp} 
\usepackage{comment}
\DeclareUnicodeCharacter{2212}{\textminus}
\usepackage{enumitem}
\usepackage{hyperref}

\newcommand{\source}{EP250916a~}

\newcommand{\xmm}{XMM-Newton}

\newcommand{\nustar}{NuSTAR~}

\begin{document}

\title{On The Nature of Einstein Probe Transient EP250916a: Insights from X-ray, Optical, and Radio Observations}

\author[0000-0002-6789-2723]{Gaurava K. Jaisawal}\email[show]{gaurava@space.dtu.dk}
\affiliation{DTU Space, Technical University of Denmark, \text{\O}rsteds Plads 348, DK-2800 Lyngby, Denmark}

\author[0000-0003-4795-7072]{Giulia Illiano}\email{}
\affiliation{INAF–Osservatorio Astronomico di Brera, Via Bianchi 46, I-23807, Merate (LC), Italy}  

\author[0000-0002-0426-3276]{Francesco Carotenuto}\email{}
\affiliation{INAF, Osservatorio Astronomico di Roma, Via Frascati 33, I-00078 Monte Porzio Catone, Italy}  

\author[]{Astrid L. Bouquin}\email{}
\affiliation{DTU Space, Technical University of Denmark, \text{\O}rsteds Plads 348, DK-2800 Lyngby, Denmark}
\affiliation{Nordic Optical Telescope, Rambla Jos\'e Ana Fern\'andez P\'erez 7, 38711 Bre\~na Baja, Spain}

\author[0000-0002-3500-631X]{David M. Russell}\email{}
\affiliation{Center for Astrophysics and Space Science (CASS), New York University Abu Dhabi, PO Box 129188, Abu Dhabi, UAE}

\author[0000-0002-8597-0756]{Giorgos Leloudas}\email{}
\affiliation{DTU Space, Technical University of Denmark, \text{\O}rsteds Plads 348, DK-2800 Lyngby, Denmark}

\author[0000-0002-0118-2649]{Andrea Sanna}\email{}
\affiliation{Dipartimento di Fisica, Universit\`a degli Studi di Cagliari, SP Monserrato-Sestu km 0.7, I-09042 Monserrato, Italy}

\author[0009-0006-4358-9929]{Dalya Akl}\email{}
\affiliation{Center for Astrophysics and Space Science (CASS), New York University Abu Dhabi, PO Box 129188, Abu Dhabi, UAE}

\author[]{Rob Fender}\email{}
\affiliation{Astrophysics, Department of Physics, University of Oxford, Keble Road, Oxford, OX1 3RH, UK}

\author[0000-0002-6154-5843]{Sara Motta}\email{}
\affiliation{INAF–Osservatorio Astronomico di Brera, Via Bianchi 46, I-23807, Merate (LC), Italy}

\begin{abstract}

We report multi-wavelength studies of the transient EP250916a, detected by the Einstein Probe on 2025 September 16. Located at low Galactic latitude, the source exhibited a rapid X-ray brightening, reaching an unabsorbed 0.5--10 keV flux of $(6.4 \pm 0.1) \times 10^{-10}$~erg~cm$^{-2}$~s$^{-1}$, followed by a plateau and a two-stage decay lasting over 40 days. Swift/XRT monitoring shows a persistently hard spectrum ($\Gamma \approx 1.6$--2.2) with only modest softening during decay, while a \nustar{} observation confirms a hard-state continuum extending up to 70~keV. Timing analysis of XMM-Newton data reveals a weak quasi-periodic oscillation (QPO) at $\sim$13~Hz. No other coherent pulsations or thermonuclear bursts are detected. Broadband spectral modeling favors a nonthermal power-law continuum with partial-covering absorption, and shows no significant thermal disk component. Optical imaging obtained with NOT/ALFOSC, LCO, and GaiaDR3 identifies two faint sources within the 2~arcsec Swift/XRT positional uncertainty.
A \textit{MeerKAT} observation at 1.28~GHz yielded no radio counterpart, with a 3$\sigma$ upper limit of 60~$\mu$Jy~beam$^{-1}$. The combination of a long-lasting outburst, a hard nonthermal X-ray spectrum, a weak QPO detection, the absence of coherent timing features, and faint potential optical counterparts disfavors a stellar-flare or extragalactic origin and supports an accreting compact-object scenario. Comparisons with similar faint, hard-state transients place EP250916a within a growing population of low-luminosity, hard-state black hole X-ray binary candidates.

\end{abstract}


\section{Introduction}

High-energy time-domain astrophysics has advanced significantly with the deployment of wide-field X-ray and $\gamma$-ray observatories, including the International Gamma-Ray Astrophysics Laboratory (INTEGRAL; \citealt{2003A&A...411L...1W}), the Neil Gehrels Swift Observatory \citep{2004ApJ...611.1005G}, the Fermi Gamma-ray Space Telescope \citep{2009ApJ...697.1071A}, the Monitor of All-sky X-ray Image (MAXI; \citealt{2009PASJ...61..999M}), Einstein Probe \citep{2022hxga.book...86Y}, and the Space-based multi-band astronomical Variable Objects Monitor (SVOM; \citealt{2022IJMPD..3130008A}). These facilities have transformed our understanding of the dynamic high-energy sky, enabling rapid discovery and coordinated follow-up of transients spanning a wide range of luminosities, timescales, and physical origins \citep[see, e.g.,][]{2004A&A...418..927R, 2012grb..book.....K, 2012arXiv1209.3114M}.

The observed population of transient X-ray sources encompasses a diverse set of astrophysical phenomena. Explosions of massive stars produce relativistic jets and afterglows identified as $\gamma$-ray bursts \citep{2012grb..book.....K, 2015JHEAp...7...73D}. Magnetically active stars can generate impulsive soft X-ray outbursts through large-scale magnetic reconnection events \citep[e.g.,][]{2012Natur.485..478M, 2021ApJ...906...72O}. In tidal disruption events, the fallback of stellar debris onto supermassive black holes powers luminous, rapidly evolving X-ray emission \citep[see, e.g.,][]{2015JHEAp...7..148K}. Accretion onto compact objects, including white dwarfs, neutron stars, and black holes, produces short to longer duration X-ray transients whose properties depend on the accretion geometry, magnetic field strength, and mass-transfer rate \citep{1997xrb..book.....L, 2007A&ARv..15....1D, 2016ApJS..222...15T}.

In binary systems, mass transfer from the companion to the compact object can form an accretion disk that emits strongly in X-rays, giving rise to an X-ray binary (XRB). XRBs rank among the brightest persistent and transient X-ray sources in the Galaxy and serve as crucial laboratories for studying accretion physics, jet formation, and relativistic effects in strong gravitational fields \citep[e.g.,][]{1997xrb..book.....L}. Their outbursts often exhibit pronounced spectral and timing evolution, reflecting changes in the structure of the inner accretion flow \citep{2007A&ARv..15....1D, 2023hxga.book..120B}. Classifying newly discovered X-ray transients, therefore, requires discriminating among multiple possible origins, many of which share overlapping X-ray characteristics. Reliable classification depends on rapid, multiwavelength follow-up, including soft-to-hard X-ray coverage and deep optical and infrared observations. This approach is particularly essential for faint or short-lived transients that might otherwise remain undetected.

The Einstein Probe (EP) mission, equipped with the Wide-field X-ray Telescope (WXT), discovered the X-ray transient EP250916a on 2025 September 16 (T$_0$=MJD~60934.148; \citealt{2025ATel17395....1D}) at low Galactic latitude ($l = 355.331^\circ$, $b = -4.634^\circ$). 
WXT monitoring revealed a rapid flux increase from $\sim 3 \times 10^{-11}$ to $\sim 2 \times 10^{-10}$~erg~cm$^{-2}$~s$^{-1}$ within $\sim 38$ hours \citep{2025GCN.41861....1D}, with a preliminary photon index of $\Gamma = 1.6 \pm 0.1$. Follow-up observations with SVOM/MXT on September 17 confirmed continued brightening, measuring a 0.2--10~keV flux of $3.5 \times 10^{-10}$~erg~cm$^{-2}$~s$^{-1}$ with $\Gamma \approx 2$ and $N_{\rm H} \approx 3 \times 10^{21}$~cm$^{-2}$ \citep{2025ATel17396....1C}. Subsequent initial Swift/XRT observations refined the source position and measured an unabsorbed 0.3--10~keV flux of $(6.2 \pm 0.1) \times 10^{-10}$~erg~cm$^{-2}$~s$^{-1}$ with $\Gamma = 1.67 \pm 0.04$ \citep{2025ATel17397....1I}. A \nustar{} observation on 2025 September 24, further revealed a stable hard-state spectrum extending up to 70~keV \citep{2025ATel17421....1J}.

To investigate the nature of \source\ and track its long-term evolution, we initiated a coordinated follow-up campaign with Swift, \nustar{}, and \xmm{}, supplemented by optical imaging with the Nordic Optical Telescope (NOT) and Las Campanas Observatory (LCO) network. We also monitored the source with \textit{MeerKAT} radio telescope. This campaign provides precise localization, continuous X-ray monitoring, and multiwavelength diagnostics, all of which are essential for constraining the physical origin of the emission from the source. 
In this paper, we present a temporal and spectral study of the X-ray outburst from EP250916a. Section~2 describes the data acquisition and reduction procedures for Swift, \nustar{}, \xmm{}, NOT, LCO, and \textit{MeerKAT} observations, including astrometric and photometric inputs from Gaia within the X-ray error box. Section~3 presents the X-ray temporal analysis, light curves, and hardness evolution. Section~4 details the spectral studies with Swift/XRT, \nustar{}, and \xmm. Section~5 discusses the implications of our results, and Section~6 summarizes our conclusions.

\begin{figure*}
\centering
\includegraphics[height=2.8in, width=3.5in, angle=0]{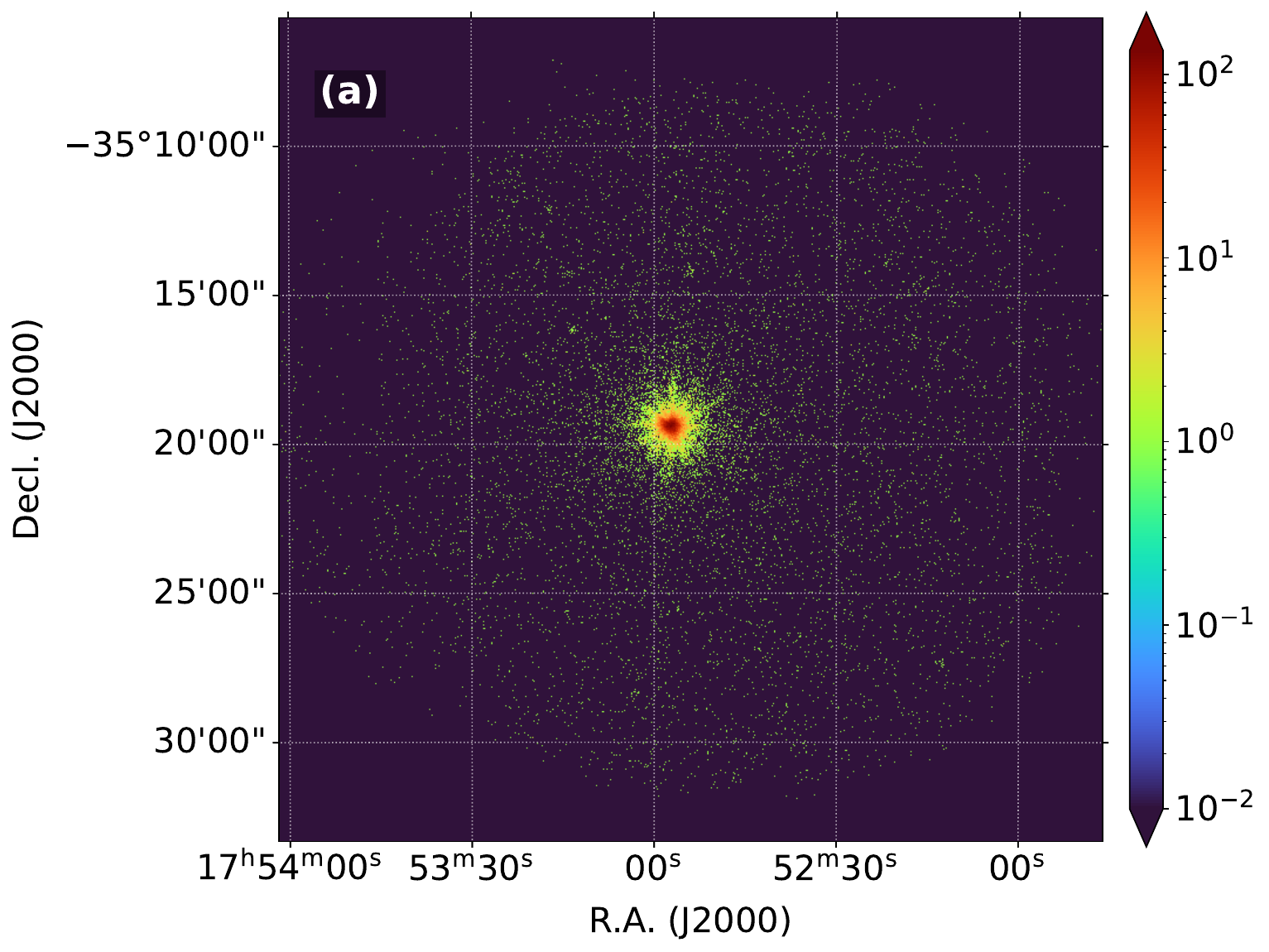}
\includegraphics[height=2.8in, width=3.5in, angle=0]{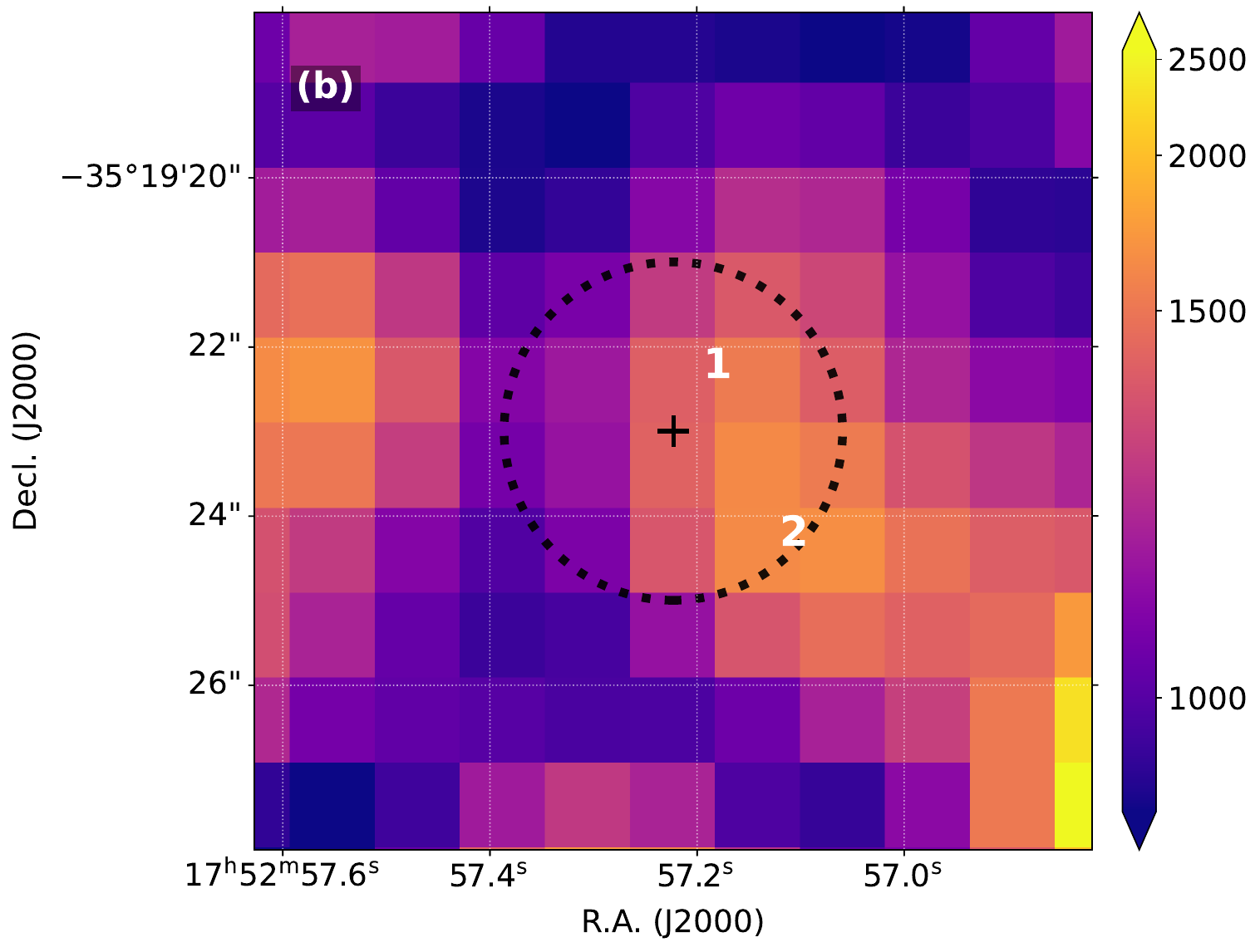} \\
\includegraphics[height=2.8in, width=3.5in, angle=0]{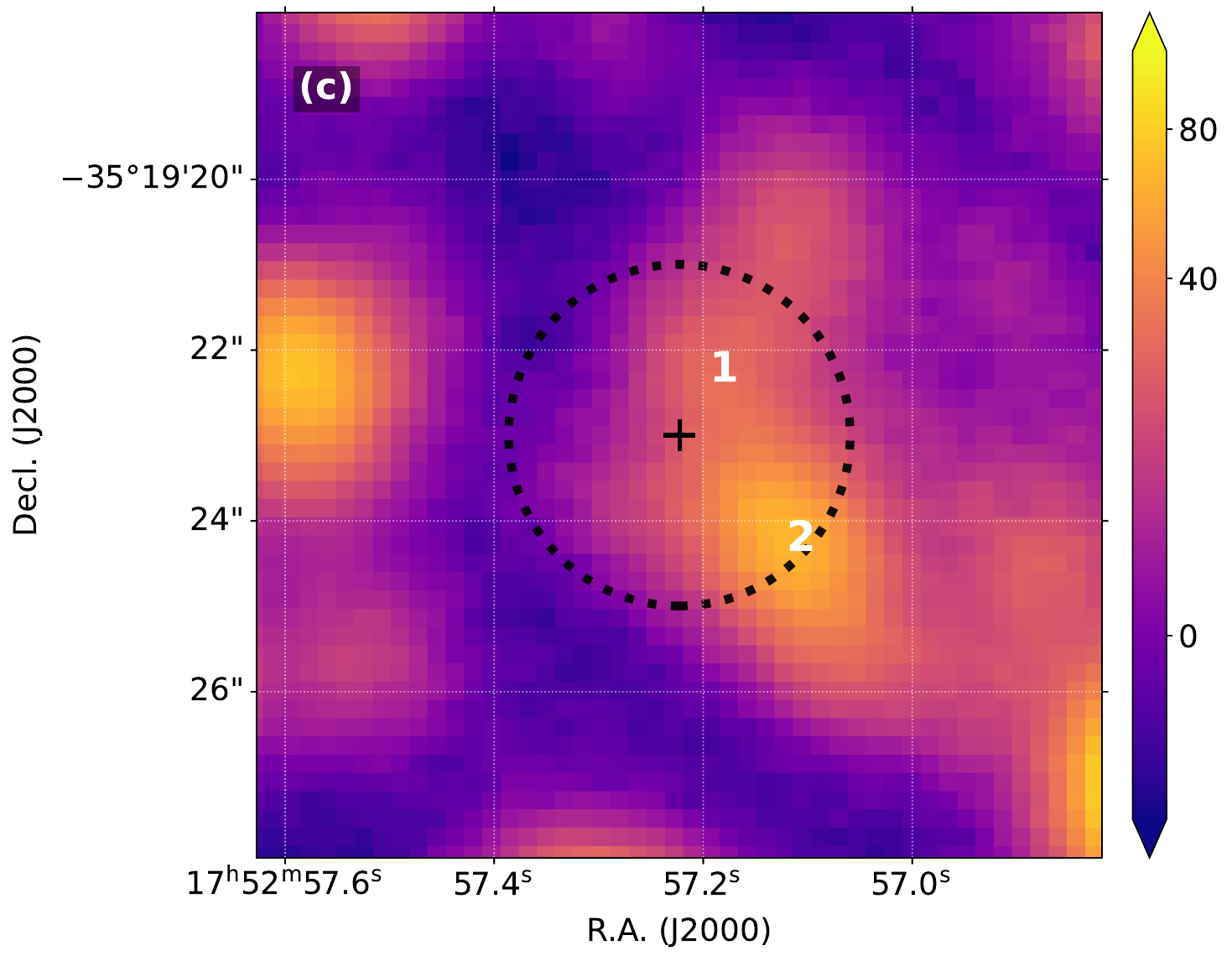} 
\includegraphics[height=2.8in, width=3.5in, angle=0]{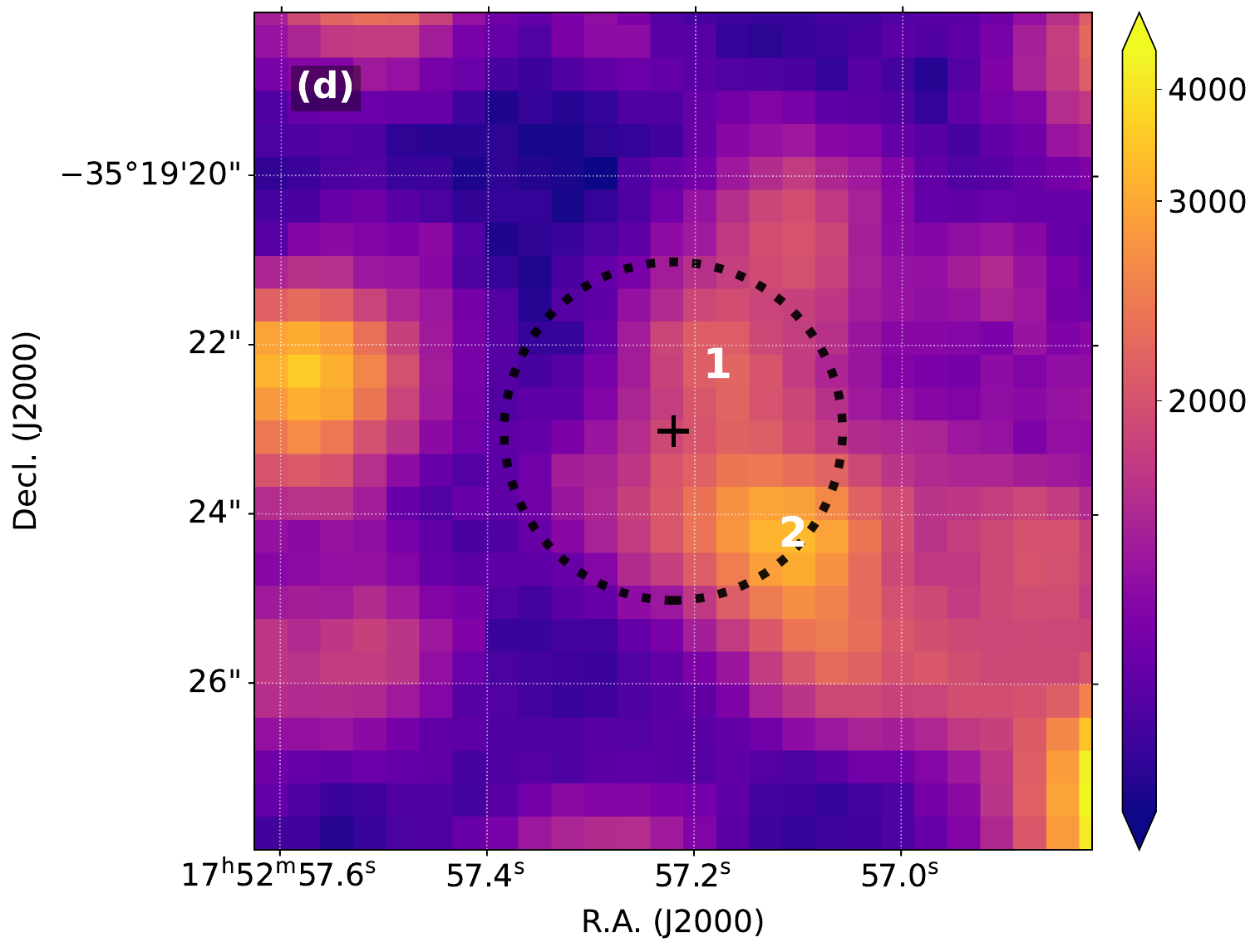} 
\caption{The field of EP250916a observed with (a) Swift/XRT (0.5--10~keV). Panels (b), (c), and (d) display zoomed views of the Swift/UVOT ($\rm V$-band), NOT/ALFOSC R-band, and LCO i$^{\prime}$-filter images, respectively, centered on the Swift/XRT position (marked by a plus sign). Each zoomed-in panel shows a $\sim8\arcsec \times 8\arcsec$ region centred on the X-ray position. Both Gaia~DR3 sources within the 2~arcsec X-ray positional uncertainty (black circle) are also marked.}
\label{fig:images}
\end{figure*}

\section{Observations and Data analysis}

\subsection{Swift}
Follow-up monitoring of EP250916a with the X-Ray Telescope (XRT) onboard the Neil Gehrels Swift Observatory \citep{2004ApJ...611.1005G} began shortly after the EP detection. 
XRT operates in two observing modes: photon counting (PC) for faint sources and windowed timing (WT) for brighter sources, allowing both spectral and timing analyses over the 0.2–10~keV energy range across a wide dynamic range. EP250916a was observed from 2025 September 17, 21:35 to November 1, 06:23 (UTC), providing 44~days of coverage (T$_0$+1.8 to T$_0$+46.3~days) with a total XRT exposure of approximately 20.6~ks. Observations were carried out simultaneously in the UV/optical bands with the Ultra-Violet/Optical Telescope (UVOT), resulting in a net UVOT exposure of 20.3~ks.
XRT data reduction was performed in \texttt{HEASoft v6.35.2} using \texttt{xrtpipeline}. Source and background spectra were extracted from both WT and PC mode events with \texttt{XSELECT}, and ancillary response files were generated using \texttt{xrtmkarf}. PC data with count rates exceeding 0.6~counts~s$^{-1}$ were corrected for pile-up by excluding the central region, with the exclusion radius determined by fitting the outer wings of the point spread function with a King function.

\source was clearly detected in the Swift/XRT observations. Its position was initially determined using standard PSF-fitting of PC mode data (total exposure 7.8~ks; panel~a of Figure~\ref{fig:images}) as R.A. (J2000) $= 268.2386^\circ$ (17$^{\rm h}$ 52$^{\rm m}$ 57.27$^{\rm s}$), Dec. (J2000) $= -35.3231^\circ$ (-35$^\circ$ 19$^{\prime}$ 23.2$^{\prime\prime}$), with a 90\% confidence error radius of 3.5~arcsec. By refining the XRT position using UVOT field astrometry, the coordinates were improved to R.A. (J2000) $= 268.23842^\circ$ (17$^{\rm h}$ 52$^{\rm m}$ 57.22$^{\rm s}$), Dec. (J2000) $= -35.32306^\circ$ (-35$^\circ$ 19$^{\prime}$ 23.0$^{\prime\prime}$), reducing the positional uncertainty to 1.9~arcsec (90\% confidence). For details on enhancing X-ray positions with UVOT astrometry, refer to \citet{2007A&A...476.1401G}. The analysis was performed using the online Swift/XRT analysis tools\footnote{\url{https://www.swift.ac.uk/user_objects/}} \citep{2009MNRAS.397.1177E}.

UVOT data were studied following standard procedures\footnote{\url{https://www.swift.ac.uk/analysis/uvot/}}. Aspect-corrected images were analyzed using a 2~arcsec source radius and a 5~arcsec background region placed in a source-free area. Multiple exposures were co-added using the \texttt{UVOTISUM} task to maximize depth. The co-added UVOT images yield AB magnitudes of $\rm B = 18.7 \pm 0.1$ and $\rm V = 18.0 \pm 0.1$~mag (see panel~b of Figure~\ref{fig:images}).

\subsection{\nustar{}}

\nustar{} \citep{2013ApJ...770..103H} is a grazing-incidence hard X-ray focusing telescope, sensitive in the 3--79~keV energy range. A \nustar{} observation of EP250916a (ObsID 91101336002; PI Jaisawal) was carried out on 2025 September 24 (MJD 60942.476--60942.927; T$_0$+8.6 days), providing a total exposure of 21~ks. The source was detected at R.A. $= 268.239^\circ$, Dec. $= -35.323^\circ$ (J2000), consistent with the localization provided by Swift/XRT. Data reduction was performed using the \texttt{HEASoft v6.35.2} package with calibration files version 20251029. Standard processing routines were followed, using \texttt{NUPIPELINE} to reprocess the unfiltered event files and \texttt{NUPRODUCTS} to extract light curves and spectra from the clean data. We performed barycentric correction using the \texttt{barycorr} task and the JPL~DE405 ephemerides.  Circular source and background regions with a radius of 60~arcsec were used in the analysis. The spectra were grouped to a minimum of 30 photons per channel bin.

\subsection{XMM-Newton}
\xmm{} \citep{2001A&A...365L...1J} observed \source on 2025 September 24 (ObsID 0973390201; PI Illiano), covering the time interval 13:44--21:09~UTC (MJD 60942.57--60942.88; T$_0$+8.6 days), yielding a total exposure of 30.3~ks. During the observation, the EPIC-pn detector \citep{2001A&A...365L..18S} was operated in fast-timing mode, providing a temporal resolution of 29.5~$\mu$s. The Observation Data Files were processed using the XMM-Newton Science Analysis Software (SAS v21.0.0). Photon arrival times were corrected to the Solar System barycenter using the \texttt{barycen} task and the JPL~DE405 ephemerides.

Intervals affected by enhanced soft-proton activity were identified from the 10–12~keV light curves and removed, leaving 26.4~ks of good exposure. Because the EPIC-pn camera provides the highest timing resolution and was unaffected by photon pile-up in fast readout modes, we used only EPIC-pn data for both the timing and spectral analysis. Source events were extracted from a 22-pixel-wide strip (RAWX~26–48), and background from a 10~pixel-wide strip (RAWX~3–13). The spectrum was grouped to a minimum of 100 counts per bin, with a 1\% systematic uncertainty added to account for instrumental features, including the Si-K line at 1.84~keV and the Au-M line at 2.2~keV  (see the official \xmm{} User Guide\footnote{\url{https://www.cosmos.esa.int/web/xmm-newton/documentation}}).

\subsection{Gaia}\label{sec:gaia}

The Gaia mission, launched by the European Space Agency in 2013, provides precise astrometric, photometric, and spectroscopic measurements for over 1.8 billion sources \citep{2023A&A...674A...1G}. Gaia observes in three primary passbands\footnote{\url{https://www.cosmos.esa.int/web/gaia/edr3-passbands}}: the broad G band (330–1050 nm) for astrometry and photometry, and the blue (G$_\mathrm{BP}$; 330–680 nm) and red (G$_\mathrm{RP}$; 640–1050 nm) bands for magnitudes and color information. 

We cross-matched the refined Swift/XRT position of EP250916a (R.A. = 268.23842$^\circ$, Dec. = -35.32306$^\circ$, 90\% confidence radius 1.9~arcsec) with the Gaia DR3 catalog. Two Gaia sources lie within 2~arcsec of the XRT position:

\begin{enumerate}[label=(\roman*)]
\item Gaia~DR3~4040690560682232064 (Source~1) at coordinates R.A. (J2000) = 268.2382413$^\circ$ and Dec. (J2000) = -35.3228499$^\circ$ is the closest object, located 0.92~arcsec separation  from the X-ray transient. It has a high-quality astrometric solution (RUWE\footnote{A goodness‑of‑fit metric that indicates how well a star’s astrometric data matches the single‑star model, see \citet{2025OJAp....8E..62E} and references therein.} of $0.942$) and a parallax $\varpi$ of $0.7945 \pm 0.2143$~mas, corresponding to a geometric distance of $D \approx 1.26$~kpc. Its photometry (G $= 18.59$~mag) indicates a faint Galactic star, with  G$_\mathrm{BP}$-G$_\mathrm{RP}$ color index of  1.3$\pm$0.1.

\item Gaia~DR3~4040690560682333568 (Source~2) at coordinates R.A. (J2000) = 268.2379343$^\circ$ and Dec. (J2000) = -35.3234010$^\circ$ lies 1.88~arcsec  from the X-ray transient. It has a noisier astrometric solution (RUWE $\approx 2.7$), likely due to unresolved binarity or source multiplicity \citep{2025OJAp....8E..62E}, with a parallax $\varpi$ of $-0.905 \pm 0.323$~mas. The object has a photometry magnitude of G~$= 17.72$~mag,  with G$_\mathrm{BP}$-G$_\mathrm{RP}$ color index of 1.37$\pm$0.03. 
 
\end{enumerate}

We also note that four additional faint Gaia sources with G=18.8--19.9 mag are located at angular separations of 2.3--3.5~arcsec, within the standard XRT positional uncertainty. However, given the refined XRT localization of 1.9~arcsec, these objects are less likely to be associated with the X-ray transient. The two Gaia sources, Source~1 and Source~2, are relatively faint, and either could plausibly be the optical counterpart to EP250916a.

\subsection{Nordic Optical Telescope}\label{sec:not}

The Nordic Optical Telescope (NOT) is a 2.56~m telescope located at the Roque de los Muchachos Observatory on La Palma, Spain. Imaging was obtained with the ALFOSC (Andalucia Faint Object Spectrograph and Camera) instrument, which provides a $6.4 \times 6.4$~arcmin$^2$ field of view and a pixel scale of 0.189~arcsec~pixel$^{-1}$. A single R-band observation of the EP250916a field was taken on 2025 October 7 at 19:58~UT (MJD 60955.83; T$_0$+21.7 days) with an exposure time of 600~s. The observation was carried out with a seeing between 1.0--1.5~arcsec with an airmass of 2.91. The data were reduced using the \texttt{PyNOT}\footnote{\url{https://github.com/jkrogager/PyNOT}} software, including standard bias subtraction and flat-field correction.
Astrometric calibration of the reduced image was performed via the online portal\footnote{\url{https://nova.astrometry.net/}} \citep{2010AJ....139.1782L}, which matches detected sources to reference catalogs.  
We calibrated the R-band photometry by cross-matching our detected sources with Gaia~DR3 stars ($\rm 10 < G < 16$) within a 0.1~deg search radius around the X-ray position of the object. The Gaia $\rm G$ magnitudes were converted to an R-band equivalent using a standard quadratic color term in  blue and red filters $\rm (G_{BP}-G_{RP})$. The photometric zero point was derived as the median difference between the transformed Gaia magnitudes and our instrumental magnitudes for the matched stars.

The R-band image is shown in panel~(c) of Figure~\ref{fig:images}. Within the 2~arcsec XRT error circle, we identify a faint optical star with R $ = 17.4 \pm 0.1$~mag (Vega), corresponding to R$_{\rm AB} \approx 17.6$~mag. This source is consistent with Gaia~DR3~4040690560682333568 (Source~2; see section~\ref{sec:gaia}). A weak emission can also be seen at the position of the first Gaia object (Source~1; Gaia DR3 4040690560682232064), with a $3\sigma$ limiting magnitude of 18.8 (Vega) in the NOT image (panel~c of Figure~\ref{fig:images}).  The reported magnitudes are not corrected for Galactic extinction.

\subsection{Las Cumbres Observatory}

We observed the field of EP250916a with the 1-m Las Cumbres Observatory (LCO) node at Cerro Tololo Inter-American Observatory, Chile, on 2025 September 19 (MJD 60937.03; T$_0$+2.9 days). 200-s exposures were taken in SDSS g$^{\prime}$ and i$^{\prime}$ filters. The airmass during observation was 1.12, and the seeing was 1.51 arcsec. Data reduction was automatically performed by the \texttt{banzai} pipeline. As with the NOT data, astrometric calibration of the reduced image was performed via the online portal. The i$^{\prime}$-band image is shown in panel~(d) of Figure~\ref{fig:images}. Both the Gaia sources (see section~\ref{sec:gaia}) within the 2~arcsec XRT error region were seen.
Photometric calibration of the $i^{\prime}$-band image was performed following the method described in Section~\ref{sec:not}. We detect Source~2 with $i^{\prime} = 17.9 \pm 0.1$~mag (Vega). A faint emission is also visible at the position of Source~1, with a $3\sigma$ limiting magnitude of 19.7~(Vega) in the LCO image (panel~d of Figure~\ref{fig:images}).

\begin{figure*}
\centering
\includegraphics[height=3in, width=3.4in, angle=0]{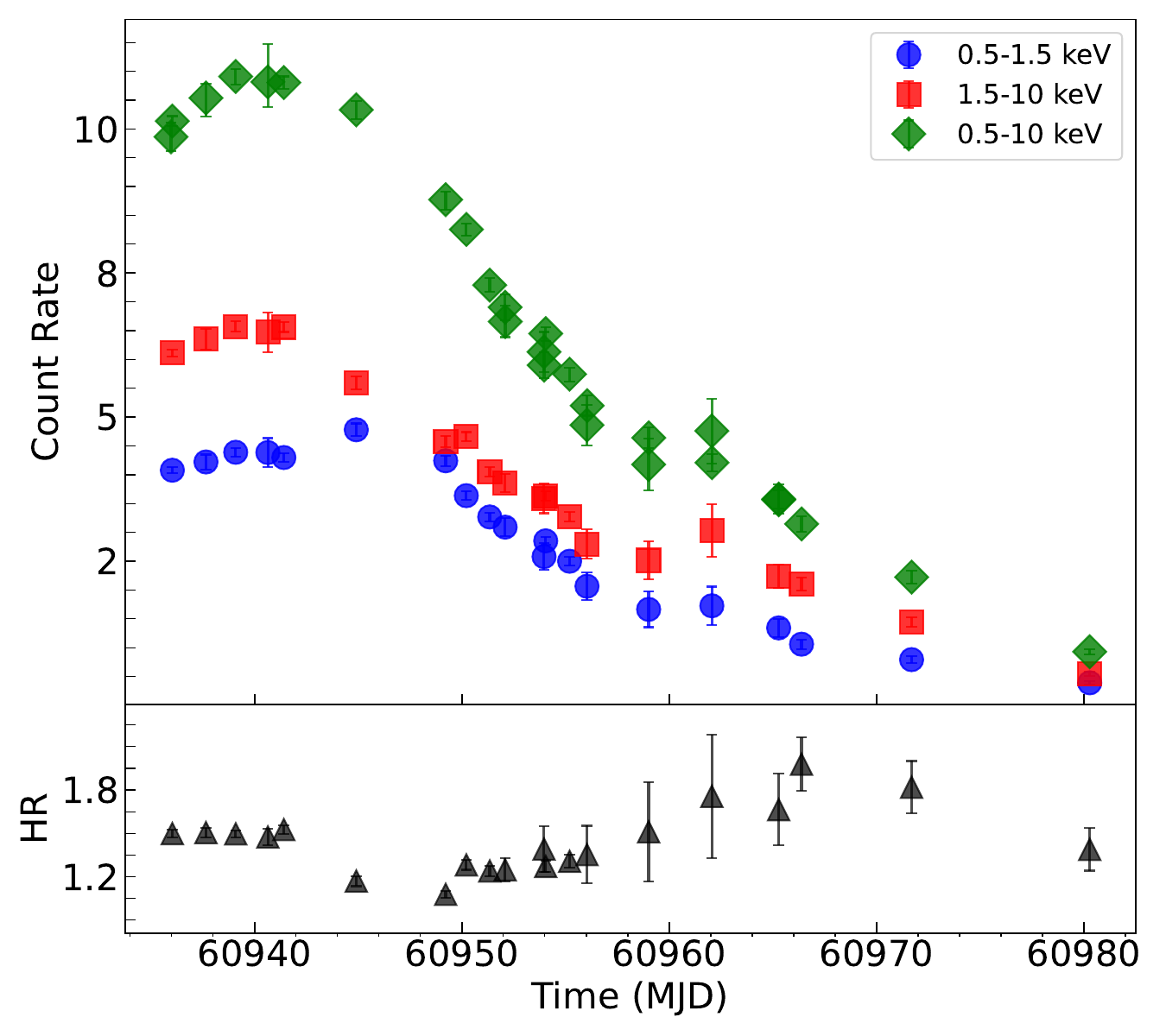} 
\includegraphics[height=3in, width=3.6in, angle=0]{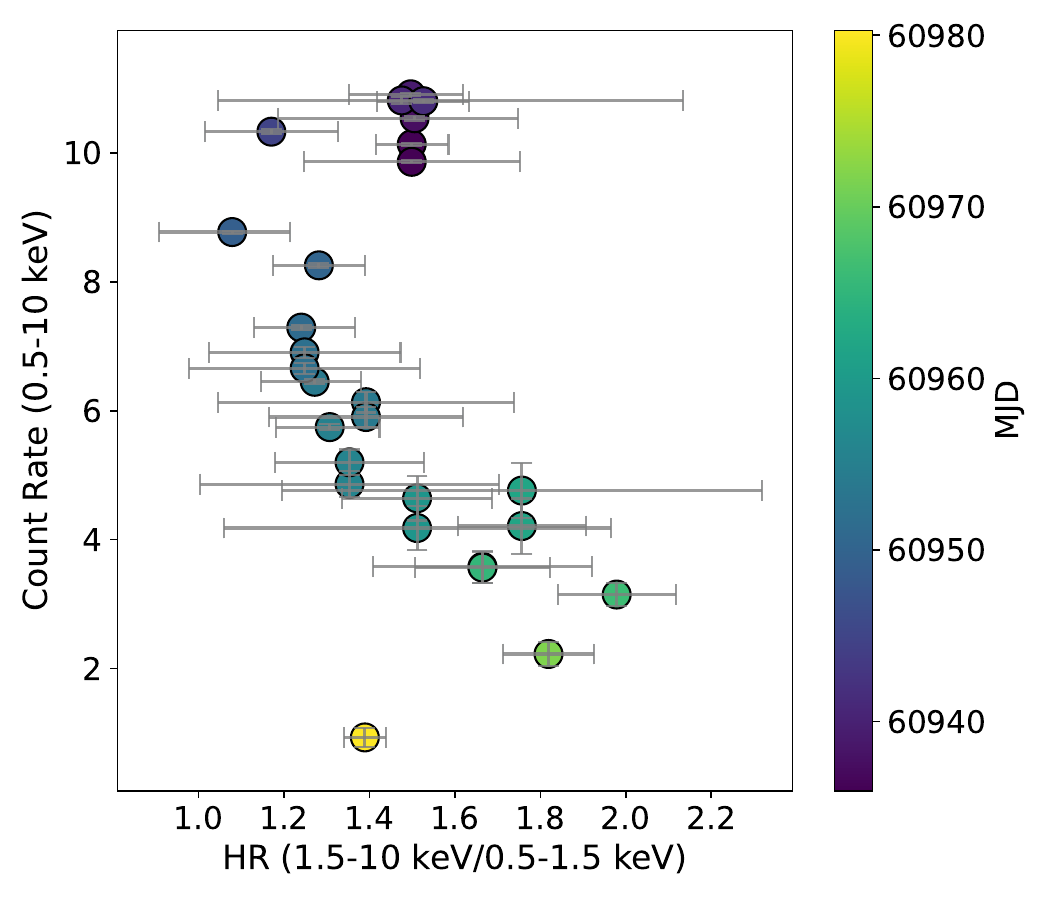}
\caption{Left: Swift/XRT light curves of EP250916a in the 0.5--1.5~keV (soft; blue circles), 1.5--10~keV (hard; red squares), and 0.5--10~keV (full; green diamonds) bands (top), with the corresponding hardness ratio (HR = 1.5--10~keV/0.5--1.5~keV) shown below. Each point represents a single Swift/XRT observation. Right: Hardness--intensity diagram illustrating the spectral evolution of EP250916a during the outburst, color-coded by time (MJD).
}  
\label{fig:xrt_lc_hr}
\end{figure*}

\subsection{MeerKAT}

We observed EP250916a with the \textit{MeerKAT} radio telescope \citep{Camilo2018, Jonas2016} as part of the X-KAT programme (SCI-20230907-RF-01; PIs Fender \& Motta), beginning on 2025 September 27 13:24 UT (MJD 60945.558; T$_0$+11.4 days) for a total on-source time of 15~minutes at L-band, with a central frequency of 1.28\,GHz and a 875\,MHz bandwidth. We used PKS~J1939-6342 as the bandpass and flux density calibrator, and PKS~1827-360 as the complex gain calibrator. We reduced the data with the \texttt{OxKAT} pipeline \citep{oxkat}, which performs standard flagging, calibration and imaging using \texttt{tricolour} \citep{Hugo_2022}, \texttt{CASA} \citep{CASA_team_2022} and \texttt{WSCLEAN} \citep{Offringa_wsclean}, respectively. For the imaging part, we adopted a Briggs weighting scheme with a $-0.3$ robust parameter, yielding a beam of $7.2\arcsec \times 5.9\arcsec$ and a $20\,\mu$Jy\,beam$^{-1}$ rms noise at 1.28 GHz. We do not detect radio emission at the known location of EP250916a, and we place a 3$\sigma$ upper limit on the peak flux density of the target at 60~$\mu$Jy beam$^{-1}$.

\section{X-ray temporal analysis}

\subsection{X-ray Light curves and hardness-intensity diagram with Swift/XRT}

To investigate the long-term temporal evolution of X-ray emission from EP250916a, we extracted Swift/XRT light curves in the soft (0.5–1.5~keV), hard (1.5–10~keV), and full (0.5–10~keV) energy bands. The light curves (top panel of Figure~\ref{fig:xrt_lc_hr}, left side) show that the source was already bright at the start of Swift monitoring and reached its observed maximum near MJD~60938. This was followed by a gradual decline, indicating that Swift captured the peak, plateau, and decay phases. The outburst lasted more than 45 days, indicating a long-lived event rather than a short flare.

The hardness ratio (HR), defined as the ratio of hard-to-soft-band count rates, provides a first look at the spectral evolution (bottom panel of Figure~\ref{fig:xrt_lc_hr}, left side). During the earliest Swift observations, the HR remained nearly constant despite moderate flux changes, consistent with a stable hard spectral state. After the peak, the HR gradually decreased, indicating mild spectral softening as the source flux declined. A secondary rise and decline in HR around MJD~60956 suggests a subtle late-time change in spectral shape, though the source remained firmly within the hard regime throughout the Swift coverage.

To further quantify the observed spectral evolution, we constructed a hardness–intensity diagram (HID; right side of Figure~\ref{fig:xrt_lc_hr}) that plots the hardness ratio against the source intensity in the 0.5--10 keV range. Because Swift did not observe the initial rising phase, the HID traces only the plateau and decay phases. During this interval, EP250916a occupies a relatively confined region of parameter space, with HR $\approx$1.2--2, exhibiting only modest changes despite a factor of $\sim$5 decrease in intensity. The HID track is smooth, showing slight softening at the highest intensities, followed by a gradual hardening (within error bars) during the decay. The absence of the rising branch in the HID prevents testing for spectral hysteresis, the characteristic ``Q-diagram'' behavior usually seen in black hole X-ray binaries that undergo full hard-to-soft transitions \citep{2005Ap&SS.300..107H}. In a typical Q-diagram, the rising branch corresponds to the initial increase in luminosity in the hard state before a transition to the soft state. In the observed interval, EP250916a traces only a single hard-branch track. The rapid rise phase that would form the lower-right leg of the Q-diagram may have occurred before Swift monitoring began. The smooth, monotonic HID evolution is consistent with the modest spectral changes seen in our spectral analysis.

\subsection{Detection of a quasi-periodic oscillation with \xmm{}}\label{sec:xmm-qpo}

We extracted the 0.5--10~keV light curve from the XMM-Newton/EPIC-pn data of \source from 2025 September 24, which shows an average count rate of $98 \pm 4$~cts~s$^{-1}$. No type-I X-ray bursts or flares were detected during the observation. We then constructed the power density spectra (PDS). The PDS was normalized such that the integral of each spectrum yields the squared fractional rms variability (power in units of rms$^2$~Hz$^{-1}$). After subtracting the white-noise contribution, the PDS was modeled with multiple Lorentzian components to characterize the broadband noise (see, e.g., \citealt{2002ApJ...572..392B}) and to search for potential pulsations or QPO-like features. 

\begin{figure}
\centering
\includegraphics[height=2.6in, width=3.35in, angle=0]{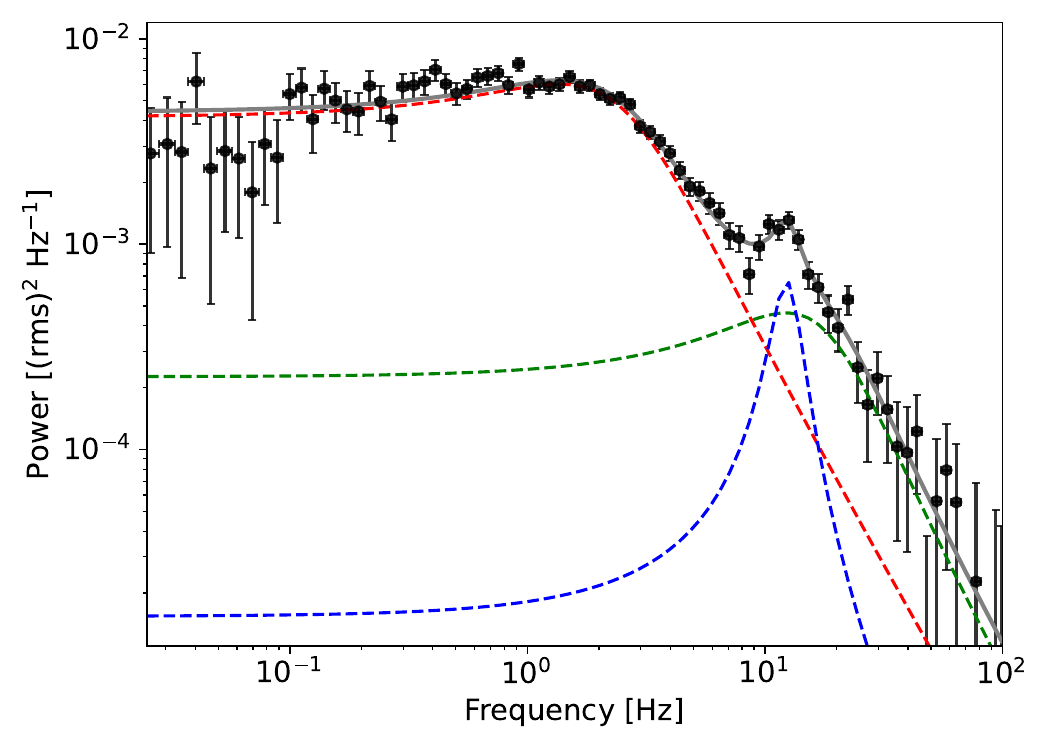}
\caption{PDS (black) of \source\ using \xmm{}. The grey line shows the best-fit model with three Lorentzian components. Broad Lorentzians (red and green) describe the underlying noise, while the narrow Lorentzian (blue) indicates a weak QPO at $\approx 13$~Hz.}
\label{fig:pds_xmm}
\end{figure}

The three Lorentzian components were required to fit the data and exhibit the following best-fit parameters with 90\% uncertainties. The first, broad component has a centroid frequency of $\nu_0 = 1.4 \pm 0.1$~Hz, a full width at half maximum of $\Delta = 4.1 \pm 0.5$~Hz, and a normalization of $\mathrm{Norm} = (6.0 \pm 0.2) \times 10^{-3}$~rms$^2$~Hz$^{-1}$. The second, broader component is centered at $\nu_0 = 12 \pm 10$~Hz, with a width of $\Delta = 24 \pm 11$~Hz and a normalization of $\mathrm{Norm} = (5 \pm 1) \times 10^{-4}$~rms$^2$~Hz$^{-1}$. These two components describe the source's underlying broadband variability. The third, narrow Lorentzian is centered at $\nu_0 = 12.3 \pm 0.7$~Hz, with a width of $\Delta = 4 \pm 3$~Hz and a normalization of $\mathrm{Norm} = (6.6 \pm 0.9) \times 10^{-4}$~rms$^2$~Hz$^{-1}$. We tentatively identified this feature as a weak QPO. We then derived a fractional rms of $7 \pm 3\%$, a characteristic frequency $\nu_{\rm max} = \sqrt{\nu_0^2 + \Delta^2} \approx 12.9 \pm 0.8$~Hz, and a quality factor $Q = \nu_0/(2\Delta) \approx 1.6 \pm 1.2$, indicating moderate coherence (see, e.g., \citealt{2002ApJ...572..392B, 2019NewAR..8501524I}). The PDS (black) and the best-fit model (grey line) are presented in Figure~\ref{fig:pds_xmm}. The contributions of the first and second broad Lorentzian components are shown in red and green, respectively, while the third, narrow Lorentzian (blue) accounts for the QPO feature at $\sim$12.3~Hz. Given its quality factor, we conservatively interpret this feature as either a Type-C-like QPO  (see, e.g., \citealt{2019NewAR..8501524I}), inner-disk oscillations, or weakly peaked noise arising from the corona of an accreting compact object.

\subsection{\nustar{} timing study}

For the \nustar{} observation on 2025 September 24,  we extracted light curves in the 3--7, 7--25, 25--50, and 3--70~keV energy bands. The full 3--70~keV light curve exhibits an average count rate of $\sim 12~\mathrm{counts~s^{-1}}$, remaining remarkably stable throughout the exposure. No short-term flares, dips, or thermonuclear bursts were detected in either energy band.
To search for coherent or quasi-periodic variability, we computed PDS from barycenter-corrected light curves using Leahy normalization, covering the frequency range $10^{-3}$–500~Hz.  
The PDS is dominated by white (Poisson) noise with an approximately flat slope. 
Confidence thresholds at $3\sigma$ and $5\sigma$ show no significant peaks, ruling out both narrow pulsations and broad QPOs during the observation. These results are consistent across both focal-plane modules and the soft, hard, and full-energy bands. 

Motivated by the tentative $\sim$13~Hz QPO detected in the XMM-Newton observation (Section~\ref{sec:xmm-qpo}), we performed a sensitivity study to constrain the upper limit on the fractional rms of a QPO at the same frequency in the \nustar{} FPMA and FPMB 3--70~keV light curves. To account for \nustar{}’s instrumental response, we performed Monte Carlo simulations incorporating the detectors’ non-paralyzable dead time. The 1~ms-binned light curves were divided into 1028~s segments, and Leahy-normalized power spectra were computed for each. Using 5000 noise-only simulations, we established the 95\% single-trial detection threshold under dead-time–modified Poisson statistics. Lorentzian QPOs with FWHM of 4~Hz (matching the \xmm{} measurement) were then injected, and 2500 simulations were performed for each trial fractional-rms amplitude across a finely sampled grid. From the simulations, we derive a 95\% confidence upper limit of 3.9\% fractional rms for a QPO at $\sim$13~Hz in the \nustar{} data.  Although this upper limit is lower than the fractional rms amplitude measured with \xmm{}, it is marginally consistent with the detection considering the uncertainties.

\section{Spectral analysis}

\subsection{Outburst Evolution with Swift/XRT}

We analyzed the Swift/XRT spectra of EP250916a in the 0.5–10~keV band using \texttt{XSPEC} \citep{1996ASPC..101...17A}. Due to low photon counts in several observations, we employed the Cash statistic (C-stat; \citealt{1979ApJ...228..939C}) for fitting. The spectra were modeled with an absorbed power-law (\texttt{tbabs*powerlaw}) using {\tt wilm} abundances \citep{2000ApJ...542..914W} and {\tt vern} cross sections \citep{1996ApJ...465..487V}. All parameters, including hydrogen column density $N_\mathrm{H}$, photon index $\Gamma$, and power-law normalization, were allowed to vary. The absorbed power-law model provided statistically acceptable fits across all epochs. We also tested single-component thermal models such as \texttt{blackbody} and \texttt{diskbb}, which yielded poorer fits, confirming that the X-ray emission is not thermally dominated. These results indicate that the soft X-ray continuum is primarily nonthermal.

\begin{figure}
\centering
\includegraphics[height=5.in, width=3.25in, angle=0]{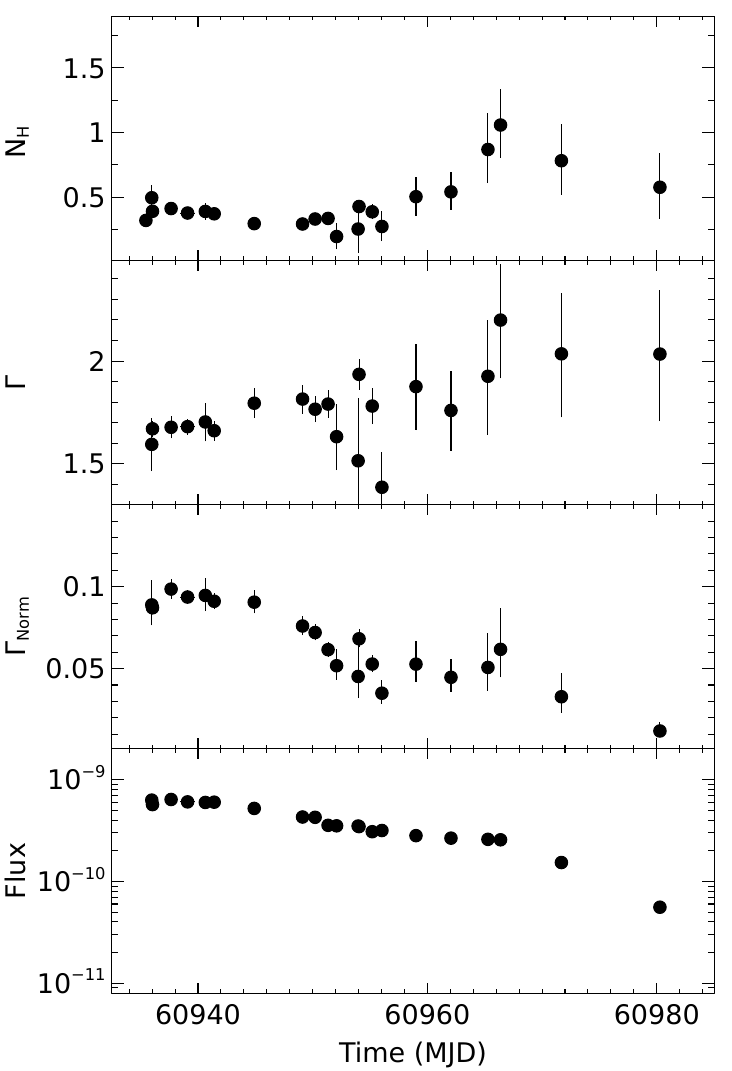}
\caption{Spectral evolution of \source\ during the outburst from Swift/XRT observations, fitted with an absorbed power-law model. Top to bottom panels show  the absorption column density $\rm N_{\rm H}$ ($10^{22}\,\mathrm{cm^{-2}}$), photon index $\Gamma$, normalization ($\Gamma_{\rm Norm}$), and the 0.5--10~keV unabsorbed flux ($\mathrm{erg\,s^{-1}\,cm^{-2}}$), respectively.}
\label{fig:xrt_spec_para}
\end{figure}

\begin{table*}
\centering
\caption{Best–fit spectral parameters for Swift/XRT observations using an absorbed power-law model. Errors are quoted at the 90\% confidence level. The 0.5-10 keV unabsorbed flux is in the unit of erg cm$^{-2}$ s$^{-1}$.}
\begin{tabular}{lccccc}
\hline
ObsIDs & N$_{\rm H}$  & $\Gamma$ & $\Gamma_{\rm Norm}$  & Flux  & Cstat (dof) \\
       & ($10^{22}$ cm$^{-2}$)  &     & ($10^{-2}$)    & ($10^{-10}$)\\
\hline
3000100001 & $0.50\pm0.10$ & $1.59\pm0.13$ & $8.9\pm1.4$ & $6.3\pm0.3$ & 94 (67) \\
3000100002 & $0.39\pm0.03$ & $1.67\pm0.04$ & $8.7\pm0.4$ & $5.7\pm0.1$ & 519 (403) \\
3000100003 & $0.41\pm0.04$ & $1.68\pm0.05$ & $9.9\pm0.6$ & $6.4\pm0.1$ & 305 (285) \\
3000100004 & $0.38\pm0.03$ & $1.68\pm0.04$ & $9.4\pm0.4$ & $6.0\pm0.1$ & 451 (384) \\
3000100005 & $0.39\pm0.07$ & $1.70\pm0.09$ & $9.5\pm1.0$ & $6.0\pm0.2$ & 128 (126) \\
3000100006 & $0.37\pm0.03$ & $1.66\pm0.05$ & $9.1\pm0.5$ & $6.0\pm0.1$ & 411 (315) \\
3000100007 & $0.30\pm0.04$ & $1.79\pm0.07$ & $9.1\pm0.7$ & $5.2\pm0.1$ & 222 (183) \\
3000100008 & $0.29\pm0.04$ & $1.81\pm0.07$ & $7.6\pm0.6$ & $4.3\pm0.1$ & 195 (188) \\
3000100009 & $0.33\pm0.04$ & $1.77\pm0.06$ & $7.2\pm0.5$ & $4.3\pm0.1$ & 243 (232) \\
3000100010 & $0.20\pm0.10$ & $1.63\pm0.16$ & $5.2\pm0.9$ & $3.5\pm0.2$ & 50 (50) \\
3000100011 & $0.34\pm0.04$ & $1.79\pm0.07$ & $6.2\pm0.5$ & $3.6\pm0.1$ & 194 (213) \\
3000100012 & $0.25\pm0.20$ & $1.51\pm0.32$ & $4.5\pm1.6$ & $3.5\pm0.5$ & 9 (14) \\
3000100013 & $0.43\pm0.05$ & $1.94\pm0.07$ & $6.8\pm0.6$ & $3.5\pm0.1$ & 234 (193) \\
3000100014 & $0.39\pm0.06$ & $1.78\pm0.09$ & $5.3\pm0.5$ & $3.1\pm0.1$ & 143 (140) \\
3000100015 & $0.27\pm0.12$ & $1.38\pm0.17$ & $3.5\pm0.7$ & $3.2\pm0.2$ & 43 (39) \\
3000100016 & $0.50\pm0.15$ & $1.88\pm0.21$ & $5.3\pm1.3$ & $2.8\pm0.2$ & 22 (31) \\
3000100017 & $0.54\pm0.15$ & $1.76\pm0.20$ & $4.5\pm1.1$ & $2.7\pm0.2$ & 35 (35) \\
3000100018 & $1.06\pm0.27$ & $2.20\pm0.28$ & $6.2\pm2.1$ & $2.6\pm0.2$ & 36 (23) \\
3000100019 & $0.87\pm0.27$ & $1.93\pm0.28$ & $5.1\pm1.8$ & $2.6\pm0.2$ & 27 (22) \\
3000100021 & $0.78\pm0.28$ & $2.03\pm0.30$ & $3.3\pm1.2$ & $1.5\pm0.1$ & 13 (19) \\
3000100022 & $0.58\pm0.25$ & $2.03\pm0.32$ & $1.2\pm0.5$ & $0.56\pm0.05$ & 20 (15) \\
\hline
\end{tabular}
\label{tab:xrt_spectral_params}
\end{table*}

The best-fit parameters for each observation are summarized in Table~\ref{tab:xrt_spectral_params}, and their temporal evolution is shown in Figure~\ref{fig:xrt_spec_para}. The absorbing column density varies between $(0.25$–$1) \times 10^{22}$~cm$^{-2}$, exceeding the Galactic line-of-sight value of $2.2\times10^{21}$~cm$^{-2}$ (HI4PI Collaboration; \citealt{2016A&A...594A.116H}). This excess may indicate intrinsic absorption local to the source, possibly associated with a dense local environment or stellar outflow material. The photon index gradually softens over time, evolving from $\Gamma \approx 1.6$ before the peak of the outburst to $\Gamma \approx 2.2$ during the late decay, suggesting a modest spectral softening. 
We also note that the hardness ratio (Figure~\ref{fig:xrt_lc_hr}) shows a slight increase, within uncertainties, during the outburst decay phase. This behavior may be caused by variations in absorption column density (Figure~\ref{fig:xrt_spec_para}), which can attenuate soft X-ray photons. As a result, the hardness ratio may show a modest increase even when the intrinsic spectrum becomes softer, as indicated by the evolution of the photon index.

Furthermore, following a rapid initial brightening to a plateau, the 0.5–10~keV unabsorbed flux exhibits a two-stage evolution during the outburst decay. This behavior suggests evolving conditions in the emission region and possible changes in the emission geometry as the outburst progresses. Incorporating rise-phase measurements from EP/WXT (\citealt{2025ATel17395....1D}) and SVOM/MXT \citep{2025ATel17396....1C}, and extrapolating them to a common 0.5–10~keV range, we modeled the Swift/XRT flux evolution together with flux from XMM-Newton/EPIC-pn\footnote{The 0.5--10~keV EPIC-pn spectrum, fitted with an absorbed power-law including a partial covering component and an emission line at 0.71~keV (see Section~\ref{sec:broadband}).}, using a quadruple broken power-law function. This model approximately captures the fast initial rise, plateau, and subsequent decay. The flux is parameterized as

{\small
\[
F(t) = 
\begin{cases}
A\left( \frac{t}{t_{\rm b1}} \right)^{+\alpha_{\rm r}} + C, &t \le t_{\rm b1} \\
A \left( \frac{t}{t_{\rm b1}} \right)^{-\alpha_{\rm p}} + C, &t_{\rm b1} < t \le t_{\rm b2} \\
A\left( \frac{t_{\rm b2}}{t_{\rm b1}} \right)^{-\alpha_{\rm p}}
\left( \frac{t}{t_{\rm b2}} \right)^{-\alpha_{1}} + C, &t_{\rm b2} < t \le t_{\rm b3} \\
A\left( \frac{t_{\rm b2}}{t_{\rm b1}} \right)^{-\alpha_{\rm p}}
\left( \frac{t_{\rm b3}}{t_{\rm b2}} \right)^{-\alpha_{1}}
\left( \frac{t}{t_{\rm b3}} \right)^{-\alpha_{2}} + C, &t > t_{\rm b3} \; ,
\end{cases}
\]
}
where $A$ is the normalization near the first break, $\alpha_{\rm r}$ the rise index, $\alpha_{\rm p}$ the plateau index, $\alpha_{1}$ and $\alpha_{2}$ the decay indices, $t_{\rm b1}$, $t_{\rm b2}$, and $t_{\rm b3}$ the break times, and $C$ a constant baseline flux.

\begin{figure}
\centering
\includegraphics[height=2.65in, width=3.4in, angle=0]{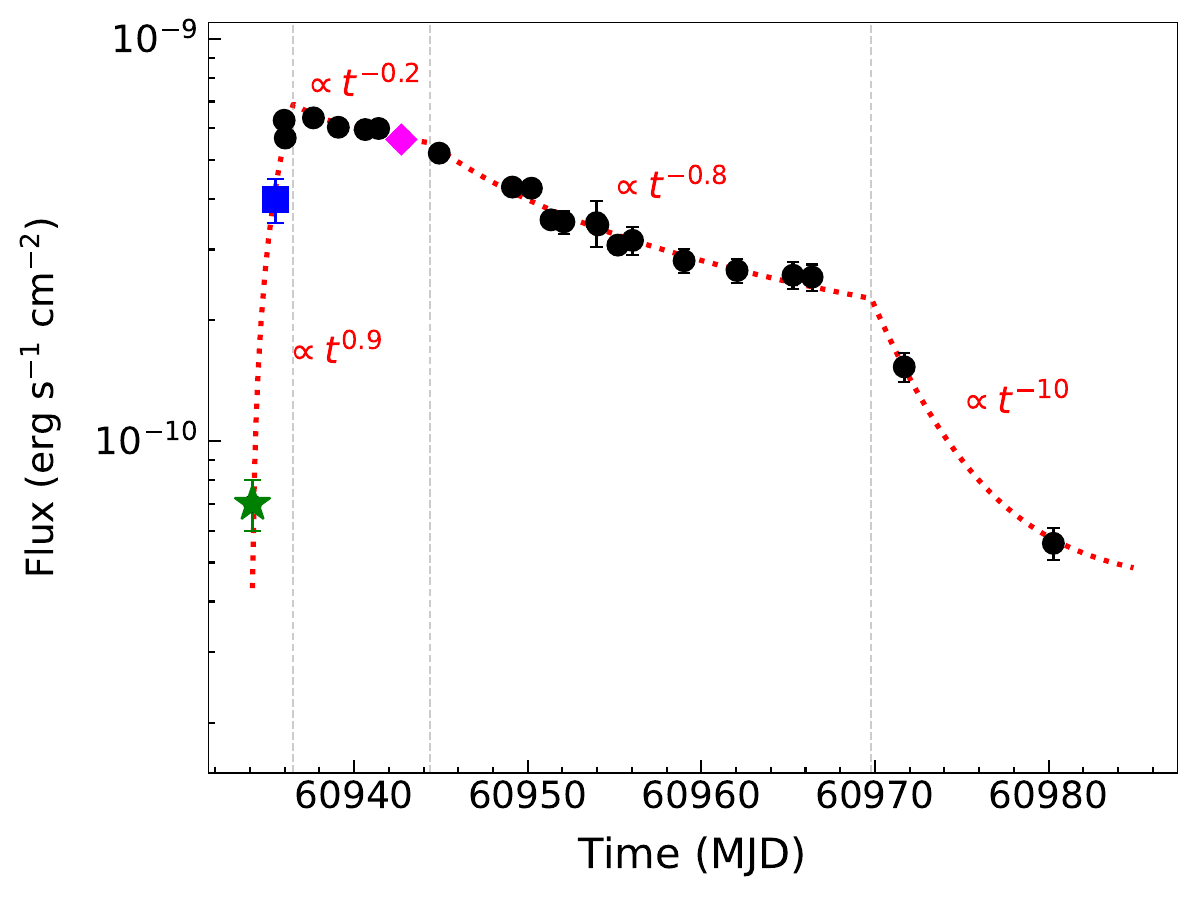}
\caption{Flux evolution of EP250916a during the outburst in the 0.5-10 keV range. EP/WXT (green star), SVOM/MXT (blue square), \xmm{} (magenta diamond), and  Swift/XRT (black circles) measurements are shown. The dotted red line shows the best-fit quadruple-broken power-law model; grey vertical lines mark the break times, with the model indices indicated.}
\label{fig:outburst-fit}
\end{figure}

The quadruple broken-power-law model approximately reproduces the overall morphology of the outburst (Figure~\ref{fig:outburst-fit}). The initial rise is steep, characterized by $\alpha_{\rm r} = 0.9 \pm 0.5$ up to $t_{\rm b1} = 2.4 \pm 0.5$~d (MJD~60936.5) since its discovery by \textit{EP}/WXT, consistent with the early rapid increase in emission. This is followed by a shallow plateau with $\alpha_{\rm p} = 0.16 \pm 0.05$ lasting until $t_{\rm b2} = 10.2 \pm 0.5$~d (MJD~60944.4), indicative of a quasi-stable emission phase. The subsequent decay proceeds with index $\alpha_{1} = 0.82 \pm 0.06$ until $t_{\rm b3} = 35.6 \pm 0.9$~d (MJD~60969.7), after which the flux drops rapidly with a steep index of $\alpha_{2} = 10 \pm 4$, pointing to an abrupt decline in the dominant emission. The normalization and constant baseline flux are $A = (6.5 \pm 0.4)\times10^{-10}$~erg~s$^{-1}$~cm$^{-2}$ and $C = (4.5 \pm 0.2)\times10^{-11}$~erg~s$^{-1}$~cm$^{-2}$, respectively. The fit yields $\chi^{2}=26$ for 15 degrees of freedom ($\chi^{2}_{\nu} = 1.73$).
The combination of progressive spectral softening and two-stage flux decay in \source is consistent with the evolution of a transient system moving from an active high-flux state toward a low-flux or quiescent phase.  Such a phenomenological approach has been used to model the outburst light curve of another X-ray transient, EP~J182730.0−095633 \citep{2025ApJ...991L..41C}, which also exhibited a fast rise followed by a two-stage decay. This modeling can be useful for understanding the temporal evolution of transient outbursts and for comparing the observed decay behavior with disk-instability–driven scenarios (e.g., \citealt{1998MNRAS.293L..42K, 2001A&A...373..251D}).

\subsection{Broadband X-ray Spectral Study with \nustar{} and \xmm{}
}\label{sec:broadband}

The 3–79~keV \nustar{} spectrum of EP250916a can be described by a simple absorbed power-law with hydrogen column density $N_\mathrm{H} = (3 \pm 2)\times10^{21}$~cm$^{-2}$, photon index $\Gamma = 2.03 \pm 0.01$ and reduced chi-squared $\chi^2_\nu = 1.16$ for 1320 degrees of freedom (Table~\ref{tab:nustar_spectral_params}). The unabsorbed flux in the 3–80~keV range is (7.01$\pm$0.01)$\times$10$^{-10}$~erg~cm$^{-2}$~s$^{-1}$.

To characterize the broadband X-ray emission, we jointly fitted simultaneous \xmm{}/EPIC-pn and \nustar{} data in the 0.5–79~keV energy range obtained on 2025 September 24. These datasets provide a comprehensive view of the broadband spectral properties during the bright, hard-state phase of the outburst. We tested multiple spectral models, including absorbed power-law, absorbed power-law plus blackbody, absorbed power-law with disk blackbody, multi-component thermal models, and Comptonization models (\texttt{comptt}). None provided satisfactory fits across the 0.5–79~keV band, primarily due to spectral residuals below $\sim 2$~keV and a strong emission line around 0.72~keV.

\begin{table*}
\centering
\caption{ Best-fitted spectral parameters of \source. The \nustar{} spectrum can be described by an absorbed power-law model. For the joint broadband \nustar{}+\xmm{} study, Model~I, Model~II, and Model~III are considered. Errors are quoted at the 90\% confidence level.}
\begin{tabular}{lcccccc}
\hline
& \multicolumn{1}{c}{\nustar{}}  & \multicolumn{3}{c}{\nustar{} + \xmm{}} \\
\cline{2-2} \cline{3-5}
Parameter                           &Power-law model      &Model~I            &Model~II                &Model~III            \\
\hline
N$_{\rm H1}$ ($10^{22}$ cm$^{-2}$)  &$0.3\pm0.2$    &$0.33\pm0.05$       &$0.30\pm0.01$        &$0.32\pm0.05$           \\
N$_{\rm H2}$ ($10^{22}$ cm$^{-2}$)   &--               &$3.9\pm0.2$          &--                       &$3.8\pm0.3$        \\
Covering Fraction                    &--              &$0.27\pm0.01$        &--                      &$0.24\pm0.01$       \\
kT$_{\rm disk}$  (keV)               &--              &--                   &$2.00\pm0.05$          &--               \\
Norm$_{\rm disk}$                     &--               &--                   &$0.24\pm0.03$          &--               \\   
$\Gamma$                            &$2.03\pm0.01$       &$2.03\pm0.01$        &$1.89\pm0.01$            &$1.99\pm0.02$     \\
$\Gamma_{\rm Norm}$                 &$0.145\pm0.003$       &$0.148\pm0.002$      &$0.094\pm0.001$          &$0.140\pm0.004$     \\
E$\rm _{cut}$ (keV)                 &--                   &--                    &--                       &$262 ^{+120} _{-64}$       \\
Energy (keV)                       &--                     &$0.71\pm0.01$          &$0.72\pm0.01$            &$0.71\pm0.01$         \\
Width (keV)                       &--                     &$0.06\pm0.01$           &$0.07\pm0.01$           &$0.06\pm0.01$         \\
Norm                               &--                    &$0.010\pm0.002$         &$0.008\pm0.001$        &$0.009\pm0.002$        \\
Flux$^*$ ($10^{-10}$)             &$8.45\pm0.04$              &$8.61\pm0.05$           &$8.49\pm0.06$           &$8.40\pm0.08$    \\
$\chi^2_\nu$ (dof)                &1.16 (1320)             &1.12 (2807)             &1.13 (2807)                       &1.11 (2806)   \\
\hline
\end{tabular}
\\
\flushleft
$*$: The 0.5-80 keV unabsorbed flux in the unit of erg~cm$^{-2}$ s$^{-1}$.\\
Note: Model~I – absorbed power-law with a partial-covering component and Gaussian line emission;  
Model~II – absorbed power-law with a disk blackbody component and Gaussian line emission;  
Model~III – absorbed cutoff power-law with a partial-covering component and Gaussian line emission.
\\
\label{tab:nustar_spectral_params}
\end{table*}

\begin{figure}
\centering
\includegraphics[height=2.8in, width=3.3in, angle=0]{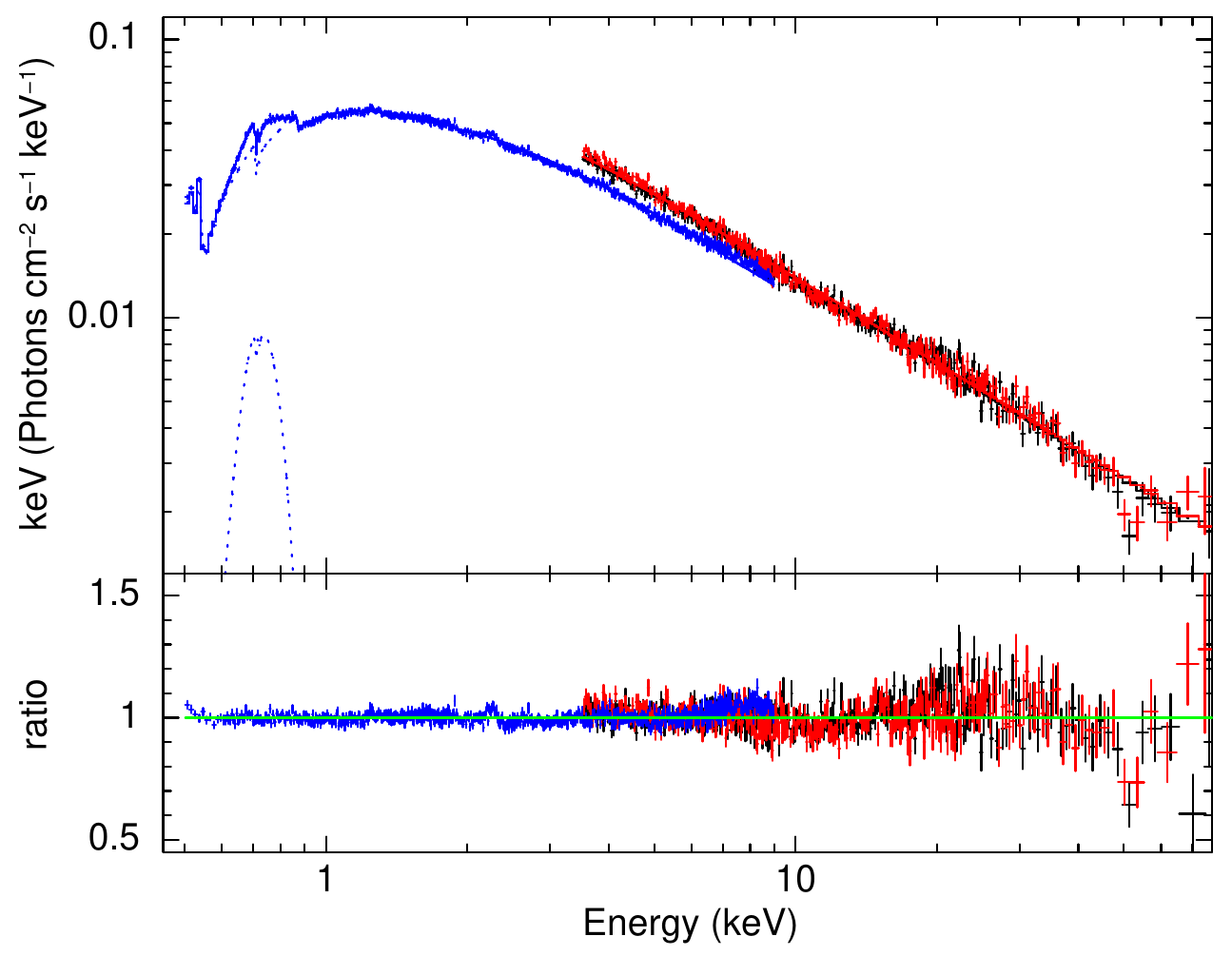}
\caption{The 0.5–79~keV broadband X-ray spectra of EP250916a from \xmm{} (blue) and \nustar{} (FPMA/FPMB; red/black) observations, fitted with an absorbed power-law model with a partial covering absorption and a Gaussian emission line (top). Spectral residuals are shown in the bottom panel.}
\label{fig:broadband_spec}
\end{figure}

A good description of the broadband continuum is obtained with an absorbed power-law modified by a partial-covering absorber, plus a Gaussian to account for the emission feature at $0.71 \pm 0.01$~keV (equivalent width $34 \pm 3$~eV). Adding the 0.71~keV line improves the fit by $\Delta\chi^{2} = 784$ for three additional degrees of freedom. Tentatively, this line may arise from a blend of highly ionized oxygen and Fe~L-shell transitions  (see, e.g. \citealt{2018Galax...6...63V, 2025ApJ...986...16J}). 
In \texttt{XSPEC} notation, the final model is
\texttt{constant $\times$ tbabs $\times$ tbpcf $\times$ (powerlaw + gaussian)} (Model~I). 
The interstellar absorption is $N_\mathrm{H} = (3.3 \pm 0.5)\times10^{21}~\mathrm{cm^{-2}}$, consistent with the Galactic line-of-sight value. The partial-covering absorber requires an additional column density of $N_{\mathrm{H2}} = (3.9 \pm 0.2)\times10^{22}~\mathrm{cm^{-2}}$ and a covering fraction of $0.27 \pm 0.01$, indicating clumpy or inhomogeneous material local to the source. The best fit yields $\chi_{\nu}^{2} \simeq 1.12$ for 2807 degrees of freedom, with a photon index of $\Gamma = 2.03 \pm 0.01$ and an unabsorbed 0.5--80~keV flux of $(8.61 \pm 0.05)\times10^{-10}~\mathrm{erg~cm^{-2}~s^{-1}}$ (see Table~\ref{tab:nustar_spectral_params}).
The broadband spectrum fitted with Model~I is shown in Figure~\ref{fig:broadband_spec}. 

As an alternative to the partial-covering component, we tested the addition of a multicolor disk blackbody (\texttt{diskbb}) to the absorbed power-law model, including the 0.7~keV Gaussian line. This model can be written as \texttt{constant $\times$ tbabs $\times$ (powerlaw + diskbb + gaussian)} (Model~II). Although Model~II provides a statistically comparable fit to Model~I, it yields a high inner disk temperature (kT$_{\rm disk} \sim$2~keV) and a low normalization (Norm$_{\rm disk} \sim 0.24$) (Table~\ref{tab:nustar_spectral_params}). 
Assuming a distance of 10~kpc, this normalization implies an apparent inner disk radius of  $\simeq 0.5$--0.8~km for typical inclination angles ($\theta \sim 0^\circ - 70^\circ$), which is smaller than expected for the innermost accretion disk radius around either a neutron star or a stellar-mass black hole. We therefore do not consider Model~II as a suitable description of the continuum and adopt Model~I as the preferred model for studying the 0.5--79 keV broadband emission from \source\ with \nustar{} and \xmm{}.

We next considered an absorbed cutoff power-law model with partial covering,
\texttt{constant $\times$ tbabs $\times$ tbpcf $\times$ (cutoffpl + gaussian)} (Model~III).
This provides a statistically comparable fit ($\chi_\nu^2 \approx 1.11$), with a photon index ($\Gamma = 1.99 \pm 0.02$) and a high-energy cutoff of $E_{\mathrm{cut}} = 262$~keV. Although the cutoff lies above the \nustar{} bandpass, the broadband data mildly prefer a turnover consistent with weak curvature at the higher energies. The spectral parameters from Models~I, II, and III are listed in Table~\ref{tab:nustar_spectral_params} (90\% confidence). No significant Fe~K$\alpha$ line or Compton reflection hump is detected in either \nustar{} or \xmm{}, suggesting an intrinsically weak reflector or a geometry in which the inner disk is truncated or weakly illuminated. The absence of any strong soft component disfavors a bright accretion disk or neutron-star boundary layer, and the lack of pulsations or thermonuclear bursts argues against a magnetized neutron star. The inferred high-energy cutoff at $\sim$260~keV, indicative of a very hot, optically thin Comptonizing corona often associated with black hole hard states, can support a nonthermal, Comptonization-dominated origin for the broadband emission, suggesting EP250916a as a hard X-ray transient.

\section{Discussion}

We present X-ray, optical, and radio studies of the newly discovered transient EP250916a, detected in 2025 September. Our X-ray observations with Swift/XRT, \nustar{}, and \xmm{} reveal a long-lived, spectrally hard X-ray transient, while complementary optical measurements from Swift/UVOT, Gaia, NOT/ALFOSC, and LCO provide additional constraints on the nature and physical origin of the source. Moreover, we obtained \textit{MeerKAT} L-band (1.28\,GHz) observation, which revealed no radio counterpart, placing a $3\sigma$ upper limit of 60~$\mu$Jy\,beam$^{-1}$. 

The X-ray outburst persisted for more than 45 days between 2025 September and November, exhibiting a rapid initial rise, a plateau, and a two-stage decay. This behavior is markedly different from stellar flares, including even the most energetic superflares. For comparison, the 40~days superflare on the K2 giant HD~251108 \citep{2024ApJ...977....6G, 2025ApJ...980..268M} reached luminosities of $\sim10^{34}$~erg~s$^{-1}$ and total radiated energies of $\sim3\times10^{39}$~erg, yet displayed soft ($kT\lesssim7$~keV) thermal plasma emission, rapid rise and decay phases, and short-lived optical brightening. EP250916a, on the other hand, shows a persistently hard X-ray spectrum, no strong high-temperature thermal component, and no rapid optical/UV brightening. Even RS~CVn-like superflares cannot account for sustained spectral hardness, multi-week decay, and faint optical/UV emission, effectively ruling out stellar coronal activity as the origin.

\subsection{Spectral Properties of \source}
To investigate the nature of this transient, we next examine the X-ray spectral characteristics indicative of accretion onto a compact object.
Swift/XRT and \nustar{} spectra of \source consistently show a hard, nonthermal power-law ($\Gamma\approx1.6$--2.2) with no significant thermal disk component, suggesting a truncated accretion disk and a persistent hard state throughout the outburst. The hardness–intensity diagram indicates that the source remains in the hard state throughout the outburst, with XRT fluxes varying by an order of magnitude from $(0.6$--$6.4)\times10^{-10}$~erg~cm$^{-2}$~s$^{-1}$, consistent with ``failed'' outbursts \citep[e.g.,][]{2016ApJS..222...15T, 2021RAA....21..125J} where the accretion rate does not reach the soft-state threshold.

At a fiducial distance of 8~kpc, approximately the Galactic Center distance, the peak XRT flux, $6.4\times10^{-10}$~erg~cm$^{-2}$~s$^{-1}$ at MJD~60937.7, yields
\begin{equation}
    L_{\rm X,peak} \approx 4.9\times10^{36}\left(\frac{d}{8~{\rm kpc}}\right)^2~{\rm erg~s^{-1}}
\end{equation}
in the 0.5--10~keV band. Similarly, the broadband unabsorbed flux measured during the \xmm{} and \nustar{} observation (MJD~60942.4--60942.9), $8.6\times10^{-10}$~erg~cm$^{-2}$~s$^{-1}$, corresponds to
\begin{equation}
    L_{\rm bol} \approx 6.6\times10^{36}\left(\frac{d}{8~{\rm kpc}}\right)^2~{\rm erg~s^{-1}}
\end{equation}
over the 0.5--80~keV energy range. These luminosities place EP250916a within the faint end of the known Galactic X-ray binary population and well below the typical peak luminosities ($\gtrsim10^{37}$~erg~s$^{-1}$) of classical bright black hole transients.

Broadband X-ray spectral modeling reveals a partial-covering absorber, indicative of clumpy or inhomogeneous material likely associated with the accretion flow. Partial-covering absorption has been observed in black-hole X-ray binaries, and can be attributed to clumpy inner accretion flows or disk winds (e.g., \citealt{2017MNRAS.468..981M, 2022MNRAS.516..124S}). At the same time, such absorption structures can appear in wind-fed high-mass X-ray binaries (HMXBs; \citealt{2015MNRAS.448..620J, 2020MNRAS.498.4830J}), but the absence of rapid, high-amplitude X-ray flaring disfavors a classical HMXB interpretation. The observed luminosity and the slow, two-stage decay favor a low-mass X-ray binary hosting either a black hole or a weakly magnetized neutron star. Moreover, the high-energy cutoff above 30~keV is not typical for accreting NSs (see, e.g., \citealt{1983ApJ...270..711W, 2019ApJ...885...18J, 2023hxga.book..147D}) and is more commonly associated with Comptonization in hot, radiatively inefficient flows around black holes \citep{2010LNP...794...53B, 2025Galax..13..111Z}.

In general, X-ray outburst profiles of low-mass X-ray binaries are known to exhibit a variety of morphologies, including fast-rise exponential-decay evolution, broken power-law decays, and multi-peaked decay structures (e.g., \citealt{1997ApJ...491..312C, 2025MNRAS.537.1146S}). The outburst lightcurve of \source (Figure~\ref{fig:outburst-fit}) is described by a quadruple broken power function, showing two step flux decay. This evolution can be understood within a disk-instability framework (e.g., \citealt{1998MNRAS.293L..42K, 2001A&A...373..251D}). In this picture, the initial rise of the X-ray outburst is associated with the propagation of a heating front through the accretion disk, while the subsequent decay reflects cooling and depletion of the accreting material. The observed change in decay slope (Figure~\ref{fig:outburst-fit}) may indicate a transition between different stages of the outburst evolution, for example from an irradiation- to a cooling-dominated phase. The two-stage flux evolution of \source therefore, broadly supports this scenario, similar to that seen in the candidate black hole binary EP~J182730.0−095633 \citep{2025ApJ...991L..41C}.

\subsection{QPO Feature}
Timing analysis reveals no significant coherent pulsations, and the \nustar{} power spectrum is consistent with noise across  10$^{-3}$--500~Hz, effectively excluding strongly magnetized neutron stars. A further diagnostic comes from our \xmm{} timing analysis, which reveals a weak QPO with a characteristic frequency of $\sim$13~Hz. Low-frequency QPOs in black hole X-ray binaries are commonly classified as Type~A, B, or C, distinguished by their coherence, amplitude, associated broadband variability, and the accretion states in which they appear \citep{2002ApJ...572..392B, 2010LNP...794...53B, 2019NewAR..8501524I, 2022MNRAS.511.3922J}. The feature detected in \source most closely resembles a Type-C–like oscillation, which is typically observed in the 0.1–30~Hz range. Its centroid frequency lies within the expected range for Type-C QPOs in intermediate states, and it is accompanied by strong, flat-topped broadband noise. This broadband variability is commonly interpreted as arising from propagating fluctuations in the mass accretion rate \citep{2002ApJ...572..392B}. Although the coherence is modest (Q$\sim 1.6$) for a Type-C QPO,  the observed centroid frequency and the presence of underlying flat-topped noise exceed that expected for Type-A QPOs and are inconsistent with the narrow $\sim$4–6~Hz range typical of Type-B~QPOs. These properties suggest that EP250916a was in, or entering, a hard-intermediate state during the \xmm{} observation on 2025 September 24 (MJD~60942; see the HID in Figure~\ref{fig:xrt_lc_hr}). In this context, the QPO may trace Lense–Thirring precession or other dynamical processes associated with a geometrically thick, precessing inner accretion flow \citep{2019NewAR..8501524I}.

\subsection{Optical Counterpart: Evidence for a Low-Mass X-ray Binary?}
Gaia~DR3 astrometry offers additional context for the optical environment of EP250916a. Within the 2~arcsec Swift/XRT error radius, two Gaia sources are detected. Source~1, located 0.92~arcsec from the X-ray centroid, has a parallax corresponding to a geometrical distance of $\sim$1.26~kpc and a magnitude of $\rm G=18.6$. Adopting the observed column density $N_{\rm H}=(3.3\pm0.5)\times10^{21}\rm\,cm^{-2}$ from our broadband spectral analysis (Table~\ref{tab:nustar_spectral_params}), and using the relation between optical extinction and hydrogen column density from \citet{Guver2009MNRAS.400.2050G}, we obtain $A_V\simeq1.5\pm0.2$~mag. Applying the empirical relation from \citet{2019ApJ...877..116W}, $A_G = 0.789\,A_V$, the corresponding extinction in the Gaia G band is $A_G\simeq1.2\pm0.2~\mathrm{mag}$. The resulting extinction-corrected absolute magnitude of Source~1, $M_G\simeq6.5$--7.2, is consistent with a mid--late K or early--mid M dwarf main-sequence star (see, e.g., \citealt{2006ima..book.....C}), possibly with a modest accretion-disk contribution.

Source~2, located 1.88~arcsec from the X-ray position, has $\rm G=17.72$~mag, broadly consistent with our NOT/ALFOSC R-band measurement, Swift/UVOT, and LCO detection. Its astrometric solution is somewhat noisy (RUWE~$\approx2.7$), possibly due to unresolved binarity or mild photocenter variability \citep{2025OJAp....8E..62E}. Assuming the same extinction ($A_G\simeq1.2\pm0.2~\mathrm{mag}$), the extinction-corrected absolute magnitude for Source~2 would be $M_G\simeq6.0$ at 1.26~kpc, $M_G\simeq3.0$ at 5~kpc, and $M_G\simeq2.0$ at 8~kpc. An intrinsically bright counterpart with $M_G\sim2$ would imply an intermediate-mass donor of $\sim$2--3~$M_\odot$ (A--F spectral type), whereas at smaller distances the photometry would instead be compatible with a late-type K--M donor (see, e.g., \citealt{2006ima..book.....C}).  Either Gaia source could be a potential counterpart, consistent with a low-mass X-ray binary scenario. 

Furthermore, LMXBs are known to show optical brightening by a few magnitudes during X-ray outbursts due to emission from the accretion disk \citep{2006MNRAS.371.1334R, 2016A&A...587A..61C}. Both candidate counterparts have magnitudes broadly consistent with Gaia measurements obtained before the X-ray outburst, without clear evidence of strong optical brightening. This weakens the case for either source being the true counterpart. Alternatively, the system may be more distant, extremely faint, and/or heavily reddened than inferred from the X-ray column density.
Future high-resolution X-ray imaging (e.g., \textit{Chandra}) together with deeper optical/infrared observations are required to confirm the true counterpart.

\subsection{X-ray and Radio Constraints}
To place EP250916a on the $L_R$--$L_X$ plane (see, e.g. \citealt{2006MNRAS.366...79M, 2013MNRAS.428.2500C}), we estimated the 5\,GHz radio luminosity using 

{\small
\begin{equation}
L_{R,5} = 4\pi d^2 \nu S_\nu \;\approx\; 3.9\times10^{26} 
\left(\frac{S_\nu}{1\,\mathrm{\mu Jy}}\right) 
\left(\frac{d}{8\,\mathrm{kpc}}\right)^2 
\mathrm{erg\,s^{-1}},
\end{equation} 
}
adopting a flat spectrum ($\alpha = 0$) to extrapolate our \textit{MeerKAT} 1.28\,GHz $3\sigma$ upper limit of 60~$\mu$Jy\,beam$^{-1}$ to 5\,GHz. This results in $L_{R,5} < 2.3 \times 10^{28}$\,erg\,s$^{-1}$ at 8\,kpc. Corresponding upper limits are $6.1\times10^{26}$\,erg\,s$^{-1}$ at 1.3 kpc, $5.2\times10^{28}$\,erg\,s$^{-1}$ at 12 kpc, and $1.45\times10^{29}$\,erg\,s$^{-1}$ at 20\,kpc.

\begin{figure}[]
\centering
\includegraphics[height=3.1in, width=3.35in, angle=0]{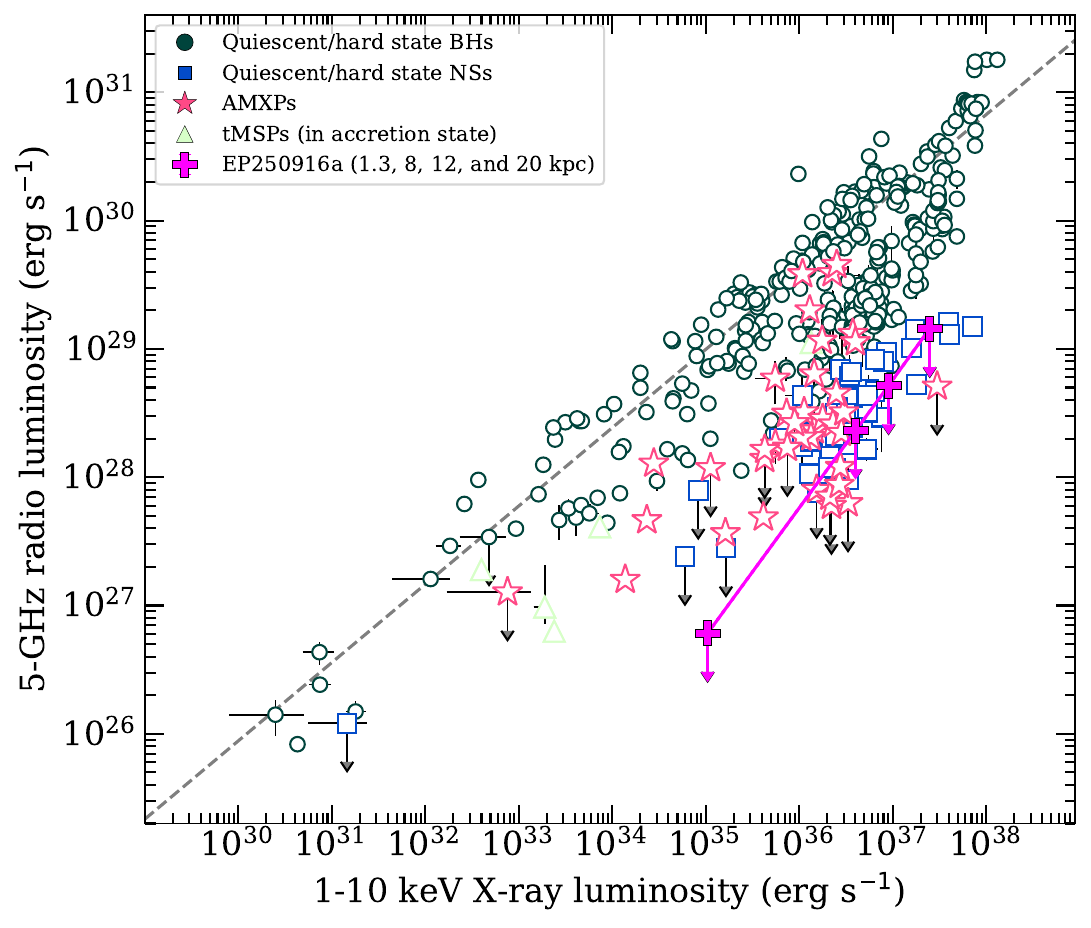}
\caption{The 1--10\,keV X-ray versus 5\,GHz radio luminosity ($L_R$--$L_X$) plane for accreting systems (see, e.g., \citealt{arash_bahramian_2022_7059313}). Gray circles represent hard-state black hole LMXBs, blue squares denote hard-state neutron star LMXBs, pink stars indicate accreting millisecond X-ray pulsars (AMXPs), and light green upward triangles correspond to transitional millisecond pulsars (tMSPs). EP250916a is shown with magenta pluses, indicating 3$\sigma$ radio upper limits at T$_0$+11.4~days, assuming distances of 1.3, 8, 12, and 20~kpc.}
\label{fig:lrlx_ep250916a}
\end{figure}

Figure~\ref{fig:lrlx_ep250916a} presents the $L_R$--$L_X$ diagram for \source. Determining the nature of the compact object in EP250916a is challenging, as there are no definitive observational signatures such as coherent pulsations, thermonuclear bursts, or dynamical mass measurements. On the 1--10\,keV X-ray versus 5\,GHz radio luminosity plane, its upper-limit points lie mostly within the region populated by hard-state neutron star systems, with no clear overlap with firmly established hard-state black hole systems. Nevertheless, the source shows a consistently nonthermal X-ray spectrum with a photon index of approximately 2, a behavior often associated with black hole accretors, leaving its classification uncertain when considered solely based on the $L_R$--$L_X$ plane.

\subsection{Nature of \source and Comparison with Faint Hard-state Transients}

EP250916a can be compared with a recently 
discovered transient EP~J182730.0--095633 (EP~J182730; \citealt{2025ApJ...991L..41C}).  Both sources exhibit persistently hard X-ray spectra with $\Gamma \approx 2$. However, their temporal evolution differs significantly. EP~J182730 experienced a short, steep outburst lasting approximately three weeks, with its X-ray luminosity declining by about three orders of magnitude. In contrast, EP250916a displays a more gradual decay, including a plateau phase, indicating a slower change in accretion rate. EP~J182730 also exhibited stronger absorption, a prominent 40~mHz QPO, and an inverted radio spectrum indicative of a compact jet, although the source was not detected in the L-band early in the outburst, similar to EP250916a. Its infrared counterpart is heavily reddened. By comparison, EP250916a shows milder absorption, only faint optical counterparts within the X-ray error region, a weaker Type-C-like low-frequency QPO around 13~Hz, and no detected radio emission during the \textit{MeerKAT} L-band observation. These contrasts illustrate the diversity among faint, hard-state X-ray transients, ranging from heavily obscured, jet-dominated systems to lower-accretion-rate objects such as EP250916a.

Historically, most black hole X-ray binaries have been discovered during bright outbursts ($\gtrsim10^{37}$ erg s$^{-1}$), reflecting the sensitivity of all-sky monitors to high-luminosity events that undergo canonical state transitions. A small subset of sources such as Swift~J1357.2−0933 \citep{2013Sci...339.1048C}, XTE~J1118+480 \citep{2001ApJ...556...42W}, XTE~J1728−295 \citep{2004ATel..229....1W}, CXOGC~J174540.0−290031 \citep{2005A&A...430L...9P}, and EP~J182730 \citep{2025ApJ...991L..41C} have been identified during intrinsically faint outbursts ($L_{\rm X} \lesssim 10^{36}$~erg~s$^{-1}$),  in addition to increasing number of very faint X-ray transients (VFXTs; see, e.g., \citealt{2006MNRAS.366L..31K}). EP250916a adds to this sparsely sampled population, illustrating the increasing ability of modern high-cadence X-ray facilities to uncover low-luminosity, hard-state accretors.

The long-duration outburst, persistently hard spectrum, faint optical counterpart, moderate luminosity, and tentative Type-C–like QPO support the interpretation of EP250916a as an accreting compact object, most likely a black hole candidate rather than a stellar flare or from an extragalactic source. Its behaviour reinforces the emerging picture of a diverse population of faint, hard-state X-ray transients revealed by contemporary time-domain surveys.

\section{Conclusions}

We have presented a comprehensive multiwavelength study of the hard X-ray transient EP250916a, discovered by the Einstein Probe on 2025 September 16. Follow-up observations with Swift/XRT, \nustar{}, \xmm{}, the NOT, LCO, and \textit{MeerKAT} provide a detailed view of its temporal, spectral, optical, and radio behavior. EP250916a exhibits a rapid initial X-ray brightening followed by a broad plateau and a two-stage decay lasting over 40 days, consistent with a sustained accretion event rather than a short-lived stellar flare. Its X-ray spectrum is persistently hard ($\Gamma \approx 1.6$–2.2), dominated by nonthermal emission. Broadband spectral modeling favors a power-law or cutoff power-law continuum modified by partial-covering absorption, indicative of clumpy or inhomogeneous material in the local environment. 

Our timing analysis with \nustar{} reveals no coherent pulsations or thermonuclear bursts, but \xmm{} detects a weak $\sim$13 Hz QPO with properties most consistent with a Type-C–like oscillation. The presence of strong broadband noise and a moderately coherent peak suggests that EP250916a was in, or entering, a hard-intermediate state during the \xmm{} observation. This QPO detection can provide evidence for dynamical processes in a geometrically thick inner accretion flow, possibly linked to Lense–Thirring precession, and strengthens the case that EP250916a is an accreting compact object rather than a stellar flare or from an extragalactic source.

The X-ray luminosity places EP250916a at the faint end of the Galactic X-ray binary population. Most transient black hole X-ray binaries have been discovered during bright outbursts ($L_X \gtrsim 10^{37}$~erg~s$^{-1}$), making systems like EP250916a relatively rare and highlighting the importance of sensitive, wide-field monitoring for identifying low-luminosity events. Comparisons with similar faint, hard X-ray transients, such as EP~J182730.0$-$095633, suggest that EP250916a may belong to an emerging population of low-luminosity, hard-state X-ray binaries or VFXTs. Variations in absorption, timing properties, and optical/infrared characteristics across these sources emphasize the diversity of accretion environments and system configurations within this class. Thus, EP250916a represents a long-duration, hard-state X-ray transient powered by accretion onto a compact object, likely a stellar-mass black hole.

\begin{acknowledgments}
We sincerely thank the referee for constructive suggestions on the paper. This work was supported by NASA through the Astrophysics Explorers Program and made use of data and software provided by the High Energy Astrophysics Science Archive Research Center (HEASARC). The data presented here were obtained with ALFOSC, which is provided by the Instituto de Astrofisica de Andalucia (IAA) under a joint
agreement with the University of Copenhagen and NOT. The MeerKAT telescope is operated by the South African Radio Astronomy Observatory, which is a facility of the National Research Foundation, an agency of the Department of Science and Innovation. This work was carried out in part using facilities and data processing pipelines developed at the Inter-University Institute for Data Intensive Astronomy (IDIA). IDIA is a partnership of the Universities of Cape Town, the Western Cape, and Pretoria.
This work received financial support from INAF through the GRAWITA 2022 Large Program Grant. DMR and DA are supported by Tamkeen under the NYU Abu Dhabi Research Institute grant CASS. GL was supported by a research grant (VIL60862) from VILLUM FONDEN. 
\end{acknowledgments}





%
\facilities{Swift(XRT and UVOT), \nustar{}, \xmm{}, Gaia, NOT:2.56m, LCO:1m,  MeerKAT}

\software{astropy
\citep{2013A&A...558A..33A,2018AJ....156..123A,2022ApJ...935..167A}, HEASoft \citep{2014ascl.soft08004N},
XSPEC \citep{1996ASPC..101...17A}, CASA \citep{CASA_team_2022}
          }



\bibliography{reference}{}
\bibliographystyle{aasjournalv7}



\end{document}